\documentclass[lettersize,journal]{IEEEtran}
\usepackage{amsmath,amsfonts}
\usepackage{array}
\usepackage[caption=false,font=scriptsize,labelfont=sf,textfont=sf]{subfig}
\usepackage{textcomp}
\usepackage{stfloats}
\usepackage{url}
\usepackage{verbatim}
\usepackage{graphicx}
\usepackage{cite}
\usepackage{multirow} 
\usepackage{booktabs} 
\usepackage{hyperref}
\usepackage[hyphenbreaks]{breakurl}
\usepackage[linesnumbered,boxed,ruled,commentsnumbered]{algorithm2e}
\usepackage{amssymb}
\usepackage{amsmath}
\usepackage{pifont}
\usepackage{xcolor}
\usepackage{subcaption}
\usepackage{bm}
\begin{document}
\newtheorem{myThe}{Theorem}
\newtheorem{myProo}{Proof}
\newtheorem{myDef}{Definition}
\newtheorem{myLem}{Lemma}
\newtheorem{myRem}{Remark}

\title{More Interactions Make Communications More Efficient:  A Directed Information Bottleneck Approach to  Task-Oriented Communication}

\author{Zhirun Zheng, Yiqin Deng, Phone Lin,~\IEEEmembership{Fellow,~IEEE}, Yuguang Fang,~\IEEEmembership{Fellow,~IEEE} 
\thanks{This work was supported in part by Hong Kong SAR Government through the Global STEM Professorship and Research Talent Hub, in part by Hong Kong Jockey Club Charities Trust under STEM Lab of Smart City (Ref.: 2023-0108).} 
\thanks{Zhirun Zheng and Yuguang Fang are with JC STEM Lab of Smart City and Department of Computer Science, City University of Hong Kong, Hong Kong, China (email: zhengzhirun2019@gmail.com; my.fang@cityu.edu.hk).}
\thanks{Yiqin Deng is with the School of Data Science, Lingnan University, Tuen Mun, Hong Kong, China (email: yiqindeng@ln.edu.hk).}
\thanks{Phone Lin is with the Department of Computer Science \& Information Engineering, National Taiwan University, Taipei, Taiwan (email: plin@csie.ntu.edu.tw).}}



\maketitle

\begin{abstract}
Task-oriented communication aims to improve communication efficiency by transmitting semantic information required to accomplish downstream tasks, yet the transmitted information may substantially overlap with previously transmitted information.
In this paper, we focus on eliminating such redundancy between current and previously transmitted information, and propose a variational directed information bottleneck joint source-channel coding (VDIB-JSCC) framework that only transmits previously unacquired information.
Specifically, we leverage directed information to measure the causal flow of task-relevant incremental information over time, and theoretically demonstrate its advantage over mutual information and conditional mutual information.
Based on this metric, we formulate a directed information bottleneck (DIB) framework that balances preserving task-relevant incremental semantic information and compressing task-irrelevant redundant information, thereby reducing communication redundancy from an information-theoretic perspective while maintaining downstream task performance.
We parameterize the encoder and decoder by utilizing the universal approximation capability of neural networks and derive a variational formulation of DIB, termed VDIB, whose tractable upper bound serves as the training loss. 
In particular, the parameterized encoder combines a Vision Transformer for frame-wise task-relevant information extraction with a Mamba-based causal state-space model to capture temporal and causal dependencies and extract incremental information, while the parameterized decoder employs a Mamba-based state-space model for task inference.
Experimental results on CIFAR-10 and CIFAR-100 based video streams demonstrate that VDIB-JSCC achieves higher task classification accuracy with less transmitted information than state-of-the-art semantic communication methods.
Our findings highlight that retaining historical knowledge enables downstream tasks to be completed using more compressed information, significantly boosting the communication efficiency in task-oriented communication systems.
\end{abstract}


\section{Introduction}
\IEEEPARstart{W}{ith} the rapid development of artificial intelligence (AI), emerging agentic AI networks~\cite{11152698,10.1145/3745238.3745531} are driving a paradigm shift in communication systems, from delivering information to humans toward facilitating intelligent decision-making by AI agents.
This paradigm shift aligns with the vision for 6G networks, in which communication is expected to support autonomous and collaborative AI agents by exchanging relevant, effective, and timely information~\cite{10597087}.
Emerging applications, including autonomous driving~\cite{10923649}, further highlight the importance of communication systems for supporting task-oriented interactions, where task relevance and decision effectiveness matter more than bit-level fidelity.
These emerging requirements challenge conventional data-oriented communication paradigms, which primarily focus on reliable data transmission, and motivate the development of task-oriented communication systems~\cite{9606667,10159007,10845817,10554663,9955312}.

In typical task-oriented communication systems, the transmitter extracts task-relevant information from source data and encodes it into compact representations, which are then transmitted over the physical channel and decoded at the receiver to support downstream tasks~\cite{9606667,10159007,10845817,10554663,9955312}.
This communication paradigm requires identifying features that are essential for task inference while discarding redundant information.
For example, in autonomous driving~\cite{10.1145/3838726,10.1145/3729420,11434993,10779389}, 
information about the positions, velocities, and intentions of surrounding vehicles and pedestrians is highly relevant to downstream tasks, such as collision avoidance and motion planning, whereas redundant visual details such as background textures and illumination variations can be discarded with negligible impact on task performance.
In particular, task-oriented communication has attracted increasing attention from the telecommunications industry due to its potential to significantly reduce communication load, with major companies such as Nokia\footnote{\url{https://www.nokia.com/}} and Ericsson\footnote{\url{https://www.ericsson.com/en}} actively exploring its key concepts and technologies~\cite{ituYSTRAfKBSC2026}.

To exploit the representation learning capability of AI for extracting task-relevant information, learning-based joint source-channel coding (JSCC) techniques~\cite{8723589,pmlrv97choi19a} have been widely explored in semantic communication to jointly optimize semantic encoding and task-oriented decoding.
However, existing studies~\cite{10845817,10597087,9606667,10159007} primarily treat each transmission independently, overlooking temporal and causal dependencies across successive transmissions and thus repeatedly transmitting previously conveyed information.
Through frequent interactions, AI agents could accumulate shared background knowledge, enabling subsequent interactions to transmit only newly acquired information. 
For instance, in autonomous driving, multiple AI agents embedded in vehicles continuously exchange environmental observations and leverage shared knowledge accumulated from prior interactions to transmit only newly emerged information, such as changes in vehicle positions or pedestrian movements, instead of repeatedly conveying previously acquired knowledge.

Furthermore, existing information-theoretic metrics~\cite{10845862,9606667,10159007,10845817,niu2024mathematicaltheorysemanticcommunication} for task-oriented communication cannot accurately quantify the task-relevant incremental information conveyed by the current transmission beyond previously acquired knowledge.
As discussed in Remark \ref{remark:directed_advantage}, mutual information~\cite{10845862,9606667,10159007} overestimates the effective information conveyed by the current transmission due to double-counting information already contained in previously acquired knowledge, whereas conditional mutual information~\cite{niu2024mathematicaltheorysemanticcommunication} underestimates it as a result of neglecting the information conveyed through causal dependencies.
Accordingly, redundant transmission of previously acquired information is an important yet largely unexplored issue in task-oriented communication: \textit{Can only incremental information be transmitted by exploiting the temporal and causal dependencies across successive communication instances, thereby further reducing communication load?}

\textbf{Our Work.} In this paper, we consider a task-oriented communication system in a temporally correlated streaming setting, where the transmitter sends an encoded representation to the receiver at each time step, and the receiver decodes the received representation to perform downstream inference tasks.
To reduce communication load arising from repeatedly transmitting information already contained in previously acquired knowledge, we propose a novel variational directed information bottleneck joint source-channel coding (VDIB-JSCC) framework that transmits only task-relevant incremental information.
Although existing approaches leverage the information bottleneck framework~\cite{tishby2000informationbottleneckmethod,9606667,10159007,10845817,10845862} to balance information compression and task relevance, they overlook the temporal and causal dependencies in sequential communication and cannot avoid repeatedly transmitting known information.
Motivated by these limitations, our work addresses the question: \textit{How can task-relevant incremental information be measured from an information-theoretic perspective, while ensuring that only the sufficient incremental information is transmitted to complete downstream task successfully?}

\textbf{Methodology.}  Unlike conventional task-oriented communication systems, we deploy knowledge bases at the transmitter and receiver to retain previously transmitted information, thereby enabling downstream tasks to be successfully completed by utilizing incremental information.
Specifically, we leverage directed information~\cite{tsur2026directedinformationestimationoptimization,6353588}, which captures temporal and causal dependencies in the sequential communication setting, to measure the task-relevant incremental information contributed by the current data sample beyond the previously acquired knowledge.
However, simply measuring incremental information does not ensure that redundant information can be removed without compromising downstream task performance.
To this end, we incorporate directed information into the information bottleneck and propose a directed information bottleneck (DIB) framework to optimize the trade-off between preserving task-relevant incremental information and compressing redundant information.
In this way, \textit{we address the fundamental question of identifying the incremental information required to accomplish downstream tasks from an information-theoretic perspective.}

Nevertheless, directly optimizing the DIB framework is intractable because it involves optimization over infinite-dimensional function spaces.
To address this challenge, we leverage the universal approximation capability of neural networks~\cite{Approximation_DNN,HORNIK1989359,HORNIK1991251} to parameterize encoder and decoder within a tractable parameterized function class.
Building on this solution strategy, we derive a variational upper bound of the DIB objective, yielding a tractable objective for neural network optimization.
We thus instantiate the encoder and decoder by combining a Vision Transformer~\cite{dosovitskiy2021imageworth16x16words,10.1145/3505244} to extract frame-wise task-relevant information with a Mamba-based causal state-space model~\cite{10.1007/978-3-031-73347-5_14} to capture temporal and causal dependencies, resulting in VDIB-JSCC, which is trained by minimizing the variational upper bound of the DIB objective.
The resulting framework progressively filters out spatial and temporal redundancy, enabling the transmission of incremental information while maintaining downstream task performance.
Furthermore, we evaluate VDIB-JSCC on the image classification datasets CIFAR-10 and CIFAR-100 against state-of-the-art task-oriented communication frameworks, demonstrating its communication efficiency while maintaining competitive downstream task performance.

\textbf{Contributions.} Our major contributions are as follows:
\begin{itemize}
    \item We propose a novel theoretical framework for task-oriented communication, termed the directed information bottleneck (DIB), which identifies task-relevant incremental information beyond previously acquired knowledge without compromising downstream task performance.
    The proposed DIB framework first utilizes directed information to measure the causal flow of task-relevant information over time and integrates it into the information bottleneck, jointly maximizing the task-relevant incremental information preserved by encodings while minimizing redundant information transmission.    
    \item Considering that the currently transmitted information could overlap with information transmitted in previous communications, directed information provides a more appropriate measure of the task-relevant effective information conveyed through the physical channel than mutual information and conditional mutual information, although its computation remains intractable.
    To obtain a tractable objective, we derive a variational upper bound of DIB utilizing the variational approximation technique and parameterize the variational distributions with neural networks, whose parameters are jointly optimized through end-to-end training.
    \item Building on the proposed DIB theoretical framework, we develop a variational directed information bottleneck joint source-channel coding (VDIB-JSCC) framework, with the transmitter extracting and encoding incremental information and the receiver decoding it for downstream tasks, assisted by the knowledge base accumulated from previous communications.
    In particular, VDIB-JSCC combines a Vision Transformer (ViT) for extracting frame-wise task-relevant information with a Mamba-based causal state-space model for capturing temporal and causal dependencies and removing redundancy already contained in the acquired knowledge.
\end{itemize}

The remainder of this paper is organized as follows. Section \ref{sec:related_works} reviews related work on task-oriented communication and information bottleneck. Section \ref{sec:sysmodel_problem} presents the system model and proposes the directed information bottleneck (DIB) framework to balance preserving task-relevant incremental information and compressing task-irrelevant redundant information. 
Section \ref{se:VDIBforSe} derives a tractable variational upper bound of the DIB objective for optimizing the encoder and decoder. Building on this upper bound, Section \ref{sec:VDIT_ISC} develops the VDIB-JSCC framework based on Vision Transformer and Mamba for task-oriented communication, enabling task completion by transmitting incremental information. Section \ref{sec:experimental} presents the experimental results, and Section \ref{sec:conclusion} concludes this paper.

\section{Related Work}\label{sec:related_works}
The information bottleneck has been widely adopted in task-oriented communication to extract task-relevant information while discarding task-irrelevant information.
We first review these concepts before presenting VDIB-JSCC.

\subsection{Task-Oriented Communication}
Recently, task-oriented communication has attracted significant attention from both academia and industry as an alternative to data-oriented communication, transmitting only task-relevant information rather than preserving bit-level fidelity to substantially reduce communication load~\cite{10554663,9955312}.
The rise of agentic AI networks has further propelled the development of semantic communication, driven by the growing demand for efficient task-oriented information exchange~\cite{10597087}.
Although the concept of semantic communication can be traced back to Shannon and Weaver's seminal work in 1971~\cite{ShannonWeaver1971}, defining task-relevant information and developing effective methods for extracting it remains challenging.
Powered by the representation learning capabilities of AI, learning-based joint source-channel coding (JSCC) techniques have been widely adopted in semantic communication to learn end-to-end mappings between source data and transmitted representations~\cite{10159007,9606667,10845817,8723589,pmlrv97choi19a}.
For example, Shao \emph{et al.}~\cite{9606667} proposed a task-oriented communication scheme that jointly optimizes feature extraction and transmission with adaptive feature sparsification to reduce communication load.
To improve robustness under practical channel conditions, Xie \emph{et al.}~\cite{10159007} studied discrete task-oriented JSCC with robust feature encoding against channel variations.
Cai \emph{et al.}~\cite{10845817} further considered cooperative edge inference and jointly optimized feature encoding, MIMO precoding, and classification under varying channel conditions.
\textit{However, existing task-oriented communication approaches generally extract task-relevant information from individual data samples, overlooking previously acquired knowledge and consequently causing redundant transmission over subsequent communication rounds.
To address this issue, VDIB-JSCC transmits only task-relevant information not already acquired from previous communications, thereby reducing redundant transmission while supporting downstream task completion.}

\subsection{Information Bottleneck}
The information bottleneck (IB)~\cite{tishby2000informationbottleneckmethod} framework formulates the trade-off between preserving task-relevant information and discarding task-irrelevant information from an information-theoretic perspective.
Given random variables $X$, $Y$, and $Z$, where $X$ denotes the source data, $Y$ represents the task target associated with $X$, and $Z$ is the task-relevant information representation extracted from $X$, the problem of learning an optimal representation $Z$ can be formulated under IB framework as
\begin{equation}
    \max_{P(z|x)} I(Z;Y) \quad \mathrm{s.t.}\quad I(X;Z) \leq I_c,
\end{equation}
where $I_c$ denotes the constraint on the amount of information retained in $Z$.
By introducing a penalty parameter $\beta$, the above constrained optimization problem could be reformulated as
\begin{equation}\label{eq:IB}
    \max_{P(z|x)} I(Z;Y)-\beta I(X;Z).
\end{equation}
Intuitively, the first term in Eq. (\ref{eq:IB}), i.e., $\max I(Z;Y)$, encourages the representation $Z$ to ``remember'' more task-relevant information associated with the target $Y$, whereas the second term, i.e., $\min I(X;Z)$, encourages $Z$ to ``forget'' task-irrelevant information from the source data $X$.
This information-theoretic principle has led to the widespread adoption of the IB framework across various fields, including AI and task-oriented communications~\cite{10438074,alemi2019deepvariationalinformationbottleneck,NIPS2015_e0040614,8253482,7133169,shwartzziv2017openingblackboxdeep,10159007,9606667,10845817}.
For instance, Alexander \emph{et al.}~\cite{alemi2019deepvariationalinformationbottleneck} developed a deep variational IB framework that derives a tractable variational approximation to the IB objective, enabling efficient optimization with neural networks through the reparameterization trick.
Tishby \emph{et al.}~\cite{7133169} further analyzed deep neural networks from an information-theoretic perspective by characterizing the evolution of mutual information between inputs, hidden representations, and output labels during training.
Goldfeld \emph{et al.}~\cite{9082644} studied representation learning from an information-theoretic perspective based on the IB framework, which balances the preservation of task-relevant information with the compression of task-irrelevant information in the input.
Building on these developments, Shao \emph{et al.}~\cite{9606667} applied the IB framework to task-oriented communication and employed a variational approximation to balance task relevance and representation compression.
\textit{Different from existing IB-based semantic communication approaches designed for independent communication settings, the proposed directed information bottleneck (DIB) framework extends the IB framework to sequential decision settings by incorporating directed information to capture causal information flow over time.}

\section{System Model and Problem Formulation}\label{sec:sysmodel_problem}
We consider an end-to-end task-oriented communication system in streaming settings, where task-relevant information is transmitted over the noisy and bandwidth-limited physical channel for downstream tasks such as image classification.
Due to the temporal correlations inherent in streaming data (e.g., video data and location trajectories), the amount of new task-relevant information not captured by previously acquired knowledge gradually decreases over time, leading to increasing information redundancy.
Before presenting the system model and problem formulation, we summarize the notation used throughout this paper in Table~\ref{tab:notations}. 
Particularly, uppercase letters (e.g., $X$) denote random variables, lowercase letters (e.g., $x$) denote their realizations, and calligraphic letters (e.g., $\mathcal{X}$) denote the corresponding realization set.
We utilize $\mathcal{X}^n$ to denote the $n$-fold Cartesian product of $\mathcal{X}$ and $|\mathcal{X}|$ to denote its cardinality.
The probability distribution of $X$, the joint distribution of $X$ and $Y$, the conditional distribution of $Y$ given $X$, and the causally conditioned distribution of $Y$ given $X$ are respectively denoted by $P_X$, $P_{X,Y}$, $P_{Y|X}$, and $P_{X \| Y}$.
Accordingly, for any realizations $x \in \mathcal{X}$ and $y \in \mathcal{Y}$, we have $P_X(x) = p(X=x)$, $P_{X,Y}(x,y) = p(X=x, Y=y)$, $P_{X|Y}(x | y) =p(X=x | Y=y)$, and $P_{X \| Y}(x \| y) = p(X = x \| Y = y)$.
Furthermore, for $i,j \in \mathbb{N}^{+}$ with $i < j$, we define the shorthand $X_i^j \triangleq (X_i, X_{i+1}, \cdots, X_j)$ and $x_i^j \triangleq (x_i, x_{i+1}, \cdots, x_j)$. For $i=1$, we simply write $X^j$ and $x^j$ by omitting the subscript, and let $X^0$ and $x^0$ denote empty sequences.

\begin{table}[h!]
	\centering
	\caption{Important notations and parameters.}\label{tab:notations}
	\begin{tabular}{cl}
		\toprule  
		\textbf{Symbol} & \textbf{Description} \\
		\midrule
        $t_k$ & Timestamp at the $k$-th time step \\
		$x_k$, $X_k$ & Sample at $t_k$ and its random variable (RV)   \\ 
        $y_k$, $Y_k$ & Target of $x_k$ and its RV \\
        $z_k$, $Z_k$ & Transmitted semantic encoding of $x_k$ and its RV\\
        $\widehat{z}_k$, $\widehat{Z}_k$ & Received semantic encoding at $t_k$ and its RV \\
        $\widehat{y}_k$, $\widehat{Y}_k$ & Inferred version of raw target $y_k$ and its RV \\
        $x^n$, $X^n$ & Sample and RV sequences up to $t_n$, $x^n  \triangleq (x_1, \cdots, x_n)$\\
        $y^n$, $Y^n$ & Target and RV sequences up to $t_n$, $y^n  \triangleq (y_1, \cdots, y_n)$  \\
        $z^n$, $Z^n$ & Raw semantic encoding and RV sequences up to $t_n$\\
        $\widehat{z}^n$, $\widehat{Z}^n$ & Noisy semantic encoding and RV sequences up to $t_n$\\
        $\widehat{y}^n$, $\widehat{Y}^n$ & Inferred target and RV sequences up to $t_n$ \\
        $D_{k}^t$, $m_{k}^t$ & Transmitter-side knowledge and memory updated at $t_k$ \\
        $D_{k}^r$, $m_k^r$ & Receiver-side knowledge and memory updated at $t_{k}$\\
        $H(\ast)$ & Entropy of random variable $\ast$ \\
        $H(\ast | \circ)$ & Conditional entropy of $\ast$ given $\circ$ \\
        $I(\star \rightarrow \ast)$ & Directed information from random variables $\star$ to $\ast$ \\
        $I(\star; \ast)$ & Mutual information between random variables $\star$ and $\ast$ \\
        $I(\star; \ast | \circ)$ & Conditional mutual information between $\star$ and $\ast$ given $\circ$  \\
        $I^{X,Z}$ & Information matrix of $\{X_k\}_{k=1}^n$ and $\{Z_k\}_{k=1}^n$ \\
        $I_{j,k}^{X,Z}$ & $(j,k)$-th entry of $I_{j,k}^{X,Z}$, $I^{X,Z} = I(X_j;Z_k | X^{j-1}, Z^{k-1})$\\
        $P(\ast \| \circ)$ & Causally conditioned distribution of $\ast$ given $\circ$ \\
        $D_{\mathrm{KL}}$ & Kullback-Leibler divergence between distributions  \\
        $\mathcal{L}_{\mathrm{DIB}}$ & Directed information bottleneck \\
        $\mathcal{L}_{\mathrm{VDIB}}$ & Variational directed information bottleneck\\
        $\mathcal{A}_{\theta}$, $\mathcal{B}_{\phi}$ & Task-oriented encoder and decoder\\
        $r_{\psi}$ &  Autoregressive prior \\
        $\beta$ & Penalty parameter balancing relevance and compression \\
        \bottomrule 
	\end{tabular}
\end{table}

\subsection{System Model}
\begin{figure}[!t]
    \centering
    \includegraphics[width=0.48\textwidth]{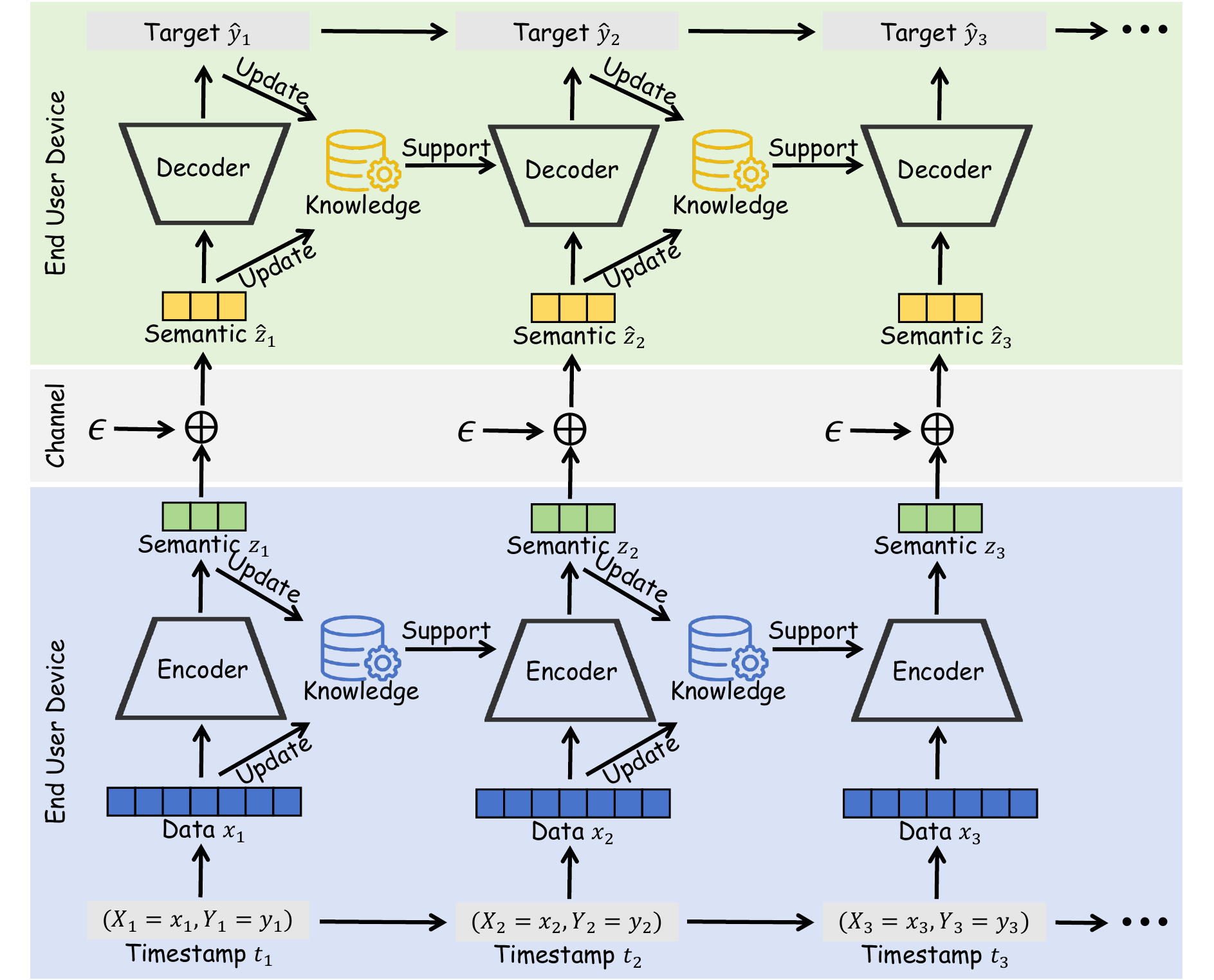}
    \caption{Diagram of the proposed task-oriented communication system for streaming settings, where temporal correlations induce information redundancy, whereas the proposed system transmits previously unacquired, task-relevant information.}
    \label{fig:semantic_systems}
\end{figure}

As depicted in Fig. \ref{fig:semantic_systems}, the considered end-to-end task-oriented communication system consists of a transmitter, a receiver, and two knowledge bases, storing historical transmitted and received information at the transmitter and receiver, respectively.
Specifically, a streaming data source generates a sequence of temporally correlated data-target pairs $\{X_k,Y_k\}_{k=1}^n$, where subscript $k$ denotes timestamp $t_k$, and the temporal correlations are characterized by the joint distribution $P(x^n, y^n) = \prod_{1 \leq k \leq n} P(x_k,y_k | x^{k-1}, y^{k-1})$.
It is worth noting that such streaming data are prevalent in real-world applications and typically exhibit strong temporal correlations, such as substantial semantic overlap between consecutive frames in a video stream~\cite{10440509}.
To mitigate semantic redundancy by leveraging temporal correlations, we introduce knowledge bases at both the transmitter and receiver, thereby enabling only the transmission of incremental task-relevant information.
In practice, these bases emulate the accumulation of shared background knowledge between the transmitter and the receiver over time, facilitating more concise semantic information during the interactive information exchanges.
Next, we present the task-oriented semantic encoding and decoding processes for streaming data, together with the transmission process over the noisy physical channel.

\begin{figure}[!t]
    \centering
    \includegraphics[width=0.50\textwidth]{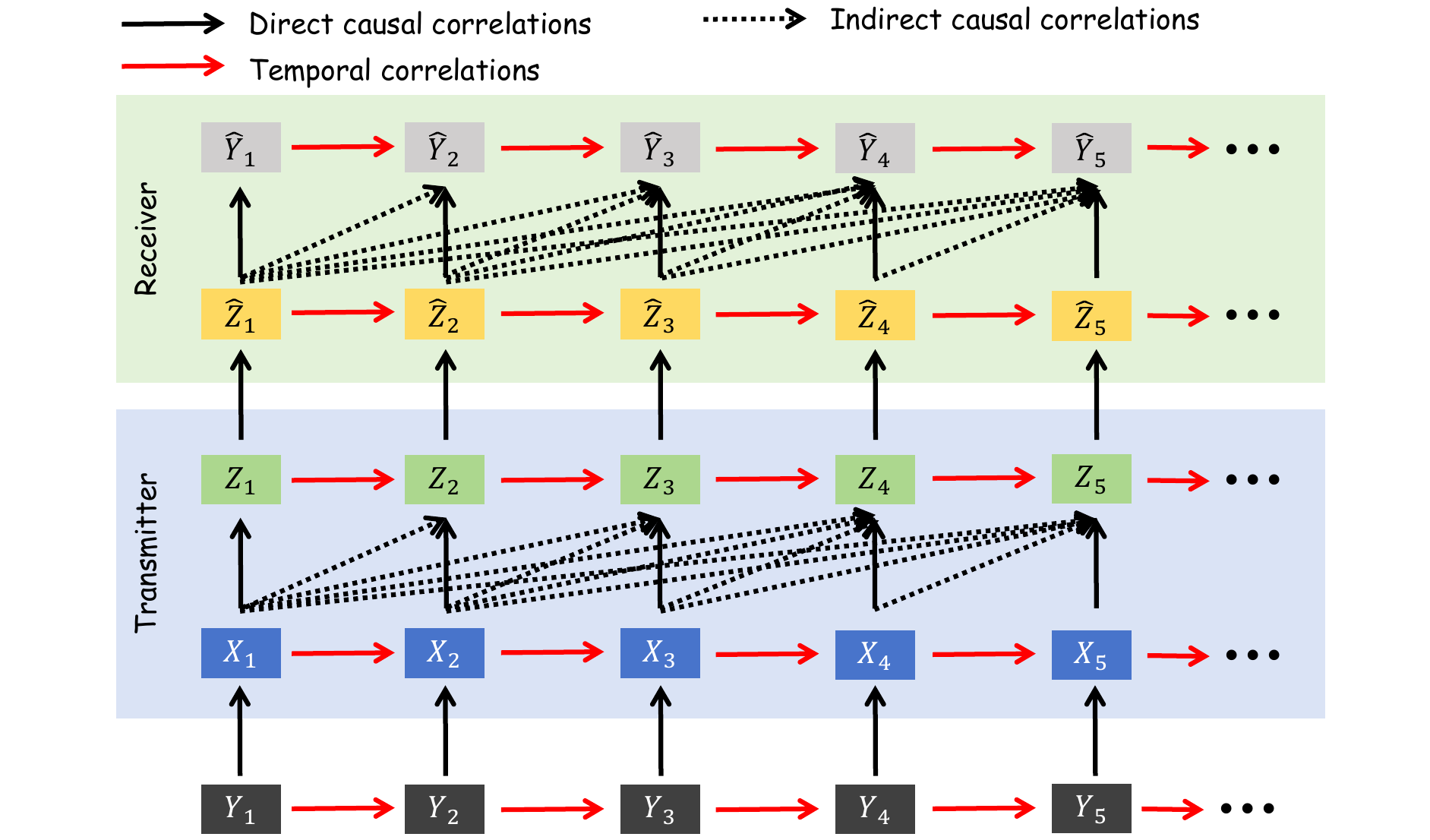}
    \caption{Relationships among the raw data stream $X^n$, transmitted semantic encoding sequence $Z^n$, received semantic encoding sequence $\widehat{Z}^n$, raw target sequence $Y^n$, and inferred target sequence $\widehat{Y}^n$. In particular, at timestamp $t_k$, the transmitter extracts the task-relevant incremental information associated with $y_k$ and encodes it into $z_k$, while the receiver infers $y_k$ from the received semantic encoding $\widehat{z}_k$, producing the inferred target $\widehat{y}_k$.}
    \label{fig:relationship}
\end{figure}

At the transmitter side, a parametric encoder $\mathcal{A}_{\theta}$ extracts the task-relevant information not captured by the knowledge base, referred to as incremental semantic information, and encodes it into a compact semantic encoding (compressed representation). 
Specifically, at timestamp $t_k$, the raw data $x_k$ is encoded into a semantic encoding $z_k$ by $\mathcal{A}_{\theta}$ with the aid of the transmitter-side knowledge base $D_{k-1}^t$, which captures the knowledge of  all previously observed raw data and semantic encodings, i.e., $D_{k-1}^t = \{x_1, \cdots, x_{k-1}\} \cup \{z_1, \cdots, z_{k-1} \}$.
Accordingly, the encoder captures the causal dependency of $z_k$ on the current and historical information $\{z^{k-1}, x^k\}$. This causal dependency implies the causally conditioned distribution of $z^n$ given $x^n$, given by\footnote{For simplicity in this paper, we assume the Markovian structure in the sequential data.} 
\begin{equation}
    P(z^n \| x^n) \triangleq \prod_{k = 1}^n P(z_k | z^{k-1},x^k),
\end{equation}
as illustrated by the light blue area in Fig. \ref{fig:relationship}.

At the physical channel side, the semantic encoding $z_k$ is transmitted over a noisy wireless channel at timestamp $t_k$, resulting in the noise-corrupted semantic encoding $\widehat{z}_k$ at the receiver.
Following existing works~\cite{9606667,10159007}, we consider an additive white Gaussian noise (AWGN) channel between the transmitter and receiver, which is widely used to model practical wireless communications.
Hence, the channel transition probability $P(\widehat{z}_k| z_k)$ is characterized by the following AWGN model
\begin{equation}\label{eq:physical_channel}
    \widehat{z}_k = z_k + \epsilon_k, ~ k = 1, 2, \cdots, n, 
\end{equation}
where $\epsilon_k$ denotes the additive channel noise sampled from a zero-mean Gaussian distribution with variance $\sigma^2$, i.e., $\epsilon_k \sim \mathcal{N}(\bm{0}, \sigma^2 \bm{I}),~ k = 1, 2, \cdots, n$.
The dependency structure between the transmitted and received semantic encoding sequences, $\widehat{Z}^n$ and $Z^n$, is depicted in the light gray area in Fig. \ref{fig:relationship}.

At the receiver side, a parametric decoder $\mathcal{B}_{\phi}$ recovers the task-relevant information from the received semantic encoding utilizing the receiver-side knowledge base.
Specifically, at timestamp $t_k$, the decoder $\mathcal{B}_{\phi}$ infers the target $\widehat{y}_k$ from the received semantic encoding $\widehat{z}_k$ utilizing the receiver-side knowledge base $D_{k-1}^r$, which contains all past received semantic encodings and inferred targets, i.e., $D_{k-1}^r = \{\widehat{z}_1, \cdots, \widehat{z}_{k-1} \} \cup \{\widehat{y}_1, \cdots, \widehat{y}_{k-1} \} $.
Thus, the decoding process induces a causal information flow from $\widehat{Z}^n$ to $\widehat{Y}^n$, where each inferred target $\widehat{y}_k$ depends causally on $(\widehat{y}^{k-1}, \widehat{z}^k)$. By further incorporating the temporal dependencies of $\widehat{Z}^n$, the causally conditioned distribution of $\widehat{y}^n$ given $\widehat{z}^n$ can be written as
\begin{equation}
    P(\widehat{y}^n \| \widehat{z}^n) \triangleq \prod_{k =1}^n P(\widehat{y}_k | \widehat{y}^{k-1}, \widehat{z}^k),
\end{equation}
representing the causal relationship between the received semantic encoding sequence $\widehat{Z}^n$ and the inferred target sequence $\widehat{Y}^n$, as shown by the light green area in Fig. \ref{fig:relationship}.

\subsection{Problem Formulation}
Intuitively,  given enough bandwidth, transmitting richer source data generally improves target inference accuracy at the receiver, but incurs higher communication load and latency over the physical channel. This reveals an inherent trade-off between inference performance and communication efficiency, which constitutes a fundamental challenge in the design of task-oriented semantic communication systems.
To reduce communication load while preserving inference accuracy, as shown in Fig.~\ref{fig:semantic_systems}, we equip both the transmitter and the receiver with knowledge bases that retain historical information, enabling the transmission of only incremental semantic information for downstream tasks.

Accordingly, the relationship among the random variables $Y^n$, $X^n$, $Z^n$, $\widehat{Z}^n$, and $\widehat{Y}^n$ is shown in Fig. \ref{fig:relationship}.
At the transmitter side, a directed causal dependence exists from $X^n$ to $Z^n$, since each semantic encoding $z_k$ is generated from the current data sample $x_k$ and the historical information $x^{k-1} \cup z^{k-1}$ (up to timestamp $t_k$) available at the encoder.
Specifically, the solid arrows represent the direct causal dependence of $z_k$ on the current raw data $x_k$ at timestamp $t_k$. In contrast, the dotted arrows represent the indirect causal dependence induced by the knowledge base, allowing past observations $x^{k-1}$ to serve as side information for encoding $z_k$.
To measure the directed information flow from a temporally correlated random sequence to another, we introduce directed information~\cite{5773045,9792384} and highlight its advantages over other information-theoretic metrics.

\begin{myDef}[Directed Information]\label{def:directed_information} 
Let $X^n \triangleq (X_1, X_2, \cdots, X_n )$ and $Z^n \triangleq (Z_1, Z_2, \cdots, Z_n)$ be two temporally correlated random sequences. The information flow from $X^n$ to $Z^n$ is measured by directed information, defined as
\begin{equation}
    I(X^n \rightarrow Z^n) = \sum_{k=1}^n I(X^k;Z_k \mid Z^{k-1}).
\end{equation}
A more intuitive interpretation of directed information $I(X^n \rightarrow Z^n)$ is the amount of information that the random sequence $(X_1, X_2, \cdots, X_n)$ causally provides about another random sequence $(Z_1, Z_2, \cdots, Z_n)$.
\end{myDef}

Furthermore, the directed information $I(X^n \rightarrow Z^n)$ can be equivalently characterized by the following entropy representation
\begin{equation}\label{eq:directed_mu}
    I(X^n \rightarrow Z^n) = \sum_{k=1}^{n} \left[ H(Z_k|Z^{k-1}) - H(Z_k|X^k,Z^{k-1}) \right].
\end{equation}
Here, the term $H(Z_k|Z^{k-1})$ measures the remaining uncertainty of $Z_k$ given the past semantic encodings $Z^{k-1}$, whereas the term $H(Z_k|X^k,Z^{k-1})$ measures the remaining uncertainty of $Z_k$ given the causally available raw data $X^k$ and the past semantic encodings $Z^{k-1}$. 
Accordingly, the difference $H(Z_k|Z^{k-1}) - H(Z_k|X^k,Z^{k-1})$ quantifies the incremental semantic information about $Z_k$ provided by the causally available raw data $X^k$ beyond that contained in the historical semantic encodings $Z^{k-1}$.
Consequently, directed information explicitly captures the temporal and causal structure of the information flow from the raw data sequence $X^n$ to the semantic encoding sequence $Z^n$.
The advantages of directed information over conventional information-theoretic metrics are highlighted in the following remark.

\begin{myRem}[Advantages of Directed Information over Mutual Information and Conditional Mutual Information]\label{remark:directed_advantage}
Let $I_{j,k}^{X,Z} = I(X_j; Z_k \mid X^{j-1}, Z^{k-1})$. We define the information matrix $I^{X,Z}$, whose $(j,k)$-th entry is given by $I_{j,k}^{X, Z}$, characterizing the temporal dependence structure between $X^n$ and $Z^n$ as
\begin{equation}\label{eq:informtation_matrix}
I^{X,Z} = 
\begin{pmatrix}
I_{1,1}^{X,Z} & I_{1,2}^{X,Z} & \cdots & I_{1,n}^{X,Z} \\
I_{2,1}^{X,Z} & I_{2,2}^{X,Z} & \cdots & I_{2,n}^{X,Z}\\
\vdots & \vdots & \ddots & \vdots \\
I_{n,1}^{X,Z} & I_{n,2}^{X,Z} & \cdots & I_{n,n}^{X,Z}
\end{pmatrix}_{n \times n.}
\end{equation}
The mutual information $I(X^n; Z^n)$ and the conditional mutual information $\sum_{j=1}^n I(X^n; Z^n | D_{i-1}^r)$ admit matrix representations based on $I^{X,Z}$, where $D_{j-1}^r$ is constructed from historical information up to timestamp $t_{j-1}$. 
Accordingly, the mutual information $I(X^n; Z^n)$ can be decomposed as
\begin{align}
    I(X^n; Z^n) &= \sum_{1 \leq j \leq k \leq n} I_{j,k}^{X,Z} + \sum_{1\leq k < j \leq n} I_{j,k}^{X,Z} \\
    &=I(X^n \rightarrow Z^n) + I(Z^{n-1} \rightarrow X^n), \label{eq:mutual_directed}
\end{align}
where the term $I(Z^{n-1} \rightarrow X^n)$ captures potential backward (non-causal) dependencies.
However, as shown in Fig. \ref{fig:relationship}, the transmitter encodes the $n$-th raw data $x_n$ into the semantic encoding $z_n$ with the aid of the knowledge base $D_{n-1}^r = \{x_1, x_2, \cdots, x_{n-1} \} \cup \{z_1, z_2, \cdots, z_{n-1} \}$, which implies that no causal influence exists from $z^{n-1}$ to $x_n$.
\textit{Consequently, the mutual information includes non-causal statistical dependencies and could overestimate the effective task-relevant information flow compared with directed information.}
In contrast, the conditional mutual information $\sum_{j=1}^n I(X_j; Z_j \mid D_{j-1}^r)$ depends only on the diagonal entries of $I^{X,Z}$ and thus ignores inter-temporal dependencies between $X^{n-1}$ and $Z^n$.
Moreover, directed information can be decomposed as
\begin{align}
I(X^n \rightarrow Z^n) &= \sum_{1\leq j \leq n} I_{j,j}^{X, Z} + \sum_{1 \leq j < k \leq n} I_{j,k}^{X,Z} \\
&=\sum_{1 \leq j \leq n} I(X_j; Z_j \mid D_{j-1}^r) + \sum_{1 \leq j < k \leq n} I_{j,k}^{X, Z}.
\end{align}
\textit{As a result, the conditional mutual information captures only instantaneous dependencies and could underestimate the effective task-relevant information flow compared with directed information.}
\end{myRem}

The encoder extracts and encodes the task-relevant incremental information, as illustrated in Fig. \ref{fig:relationship}, thereby minimizing communication load while preserving inference performance of downstream tasks.
To this end, we propose a novel directed information bottleneck (DIB) framework for semantic communications, formulated as the following optimization problem
\begin{equation}\label{eq:DIB}
  \max_{P(\widehat{z}^n \| x^n)}  \frac{1}{n} \big[\underbrace{ I(Y^n \rightarrow \widehat{Z}^n)}_{\text{Semantic Relevance}} - \beta \underbrace{I(X^n \rightarrow \widehat{Z}^n)}_{\text{Compression Cost}} \big].
\end{equation}
The first term in Eq. (\ref{eq:DIB}), i.e., $I(Y^n \rightarrow \widehat{Z}^n)$, quantifies the causal relevance of the compressed representation $\widehat{Z}^n$ to the target sequence $Y^n$.
The second term, i.e., $I(X^n \rightarrow \widehat{Z}^n)$, measures the total information retained from the raw data sequence $X^n$, including both task-relevant and task-irrelevant information, and thus serves as a compression cost.
In particular, the difference $I(X^n \rightarrow \widehat{Z}^n) - I(Y^n \rightarrow \widehat{Z}^n)$ measures the causal semantic redundancy, i.e., the portion of retained information that is irrelevant to the downstream task.
Note that the decision variable in Eq. (\ref{eq:DIB}) is the causally conditioned distribution of $\widehat{Z}^n$ given $X^n$, denoted by
\begin{equation}
    P(\widehat{z}^n \| x^n) \triangleq \prod_{k=1}^n P(\widehat{z}_k | \widehat{z}^{k-1},x^k).
\end{equation}

The relations shown in Fig. \ref{fig:relationship} imply the data processing inequality, which characterizes the relationship between the semantic relevance and compression cost terms in Eq. (\ref{eq:DIB}). The explicit form is given in Theorem \ref{them:data_inequality}.

\begin{myThe}\label{them:data_inequality}
Let the temporally correlated random sequences $Y^n$, $X^n$, $Z^n$, and $\widehat{Z}^n$ satisfy the causal relations illustrated in Fig. \ref{fig:relationship}. Then, the following data processing inequalities hold
\begin{equation}\label{eq:data_proc}
\begin{aligned}
    I(Y^n \rightarrow \widehat{Z}^n) \leq I(X^n \rightarrow \widehat{Z}^n) & \leq \sum_{k=1}^n I(\widehat{Z}_k; Z_k | \widehat{Z}^{k-1}) \\
    & \leq \sum_{k=1}^n I(\widehat{Z}_k;Z_k) \\
    &\leq n\cdot C,
\end{aligned}
\end{equation}
where $C$ denotes the Shannon channel capacity of the extended channel, given by
\begin{equation}
C = \max_{k \in [1,n]}  \max_{P(z_k)} I(\widehat{Z}_k;Z_k).
\end{equation}
In particular, directed information quantifies the total causal information flow from one random sequence to another.
Dividing both sides of Eq. \eqref{eq:data_proc} by $n$ yields the following data processing inequality for the average information flow
\begin{equation}\label{eq:them_2}
    \frac{1}{n}I(Y^n \rightarrow \widehat{Z}^n) \leq \frac{1}{n}I(X^n \rightarrow \widehat{Z}^n) \leq \max_{k \in [1,n] } I(\widehat{Z}_k; Z_k) \leq C.
\end{equation}
\end{myThe}

\begin{myProo}
The proof is given in Appendix \ref{app:theorem_1}.
\end{myProo}

\begin{figure}[!t]
    \centering
    \includegraphics[width=0.50\textwidth]{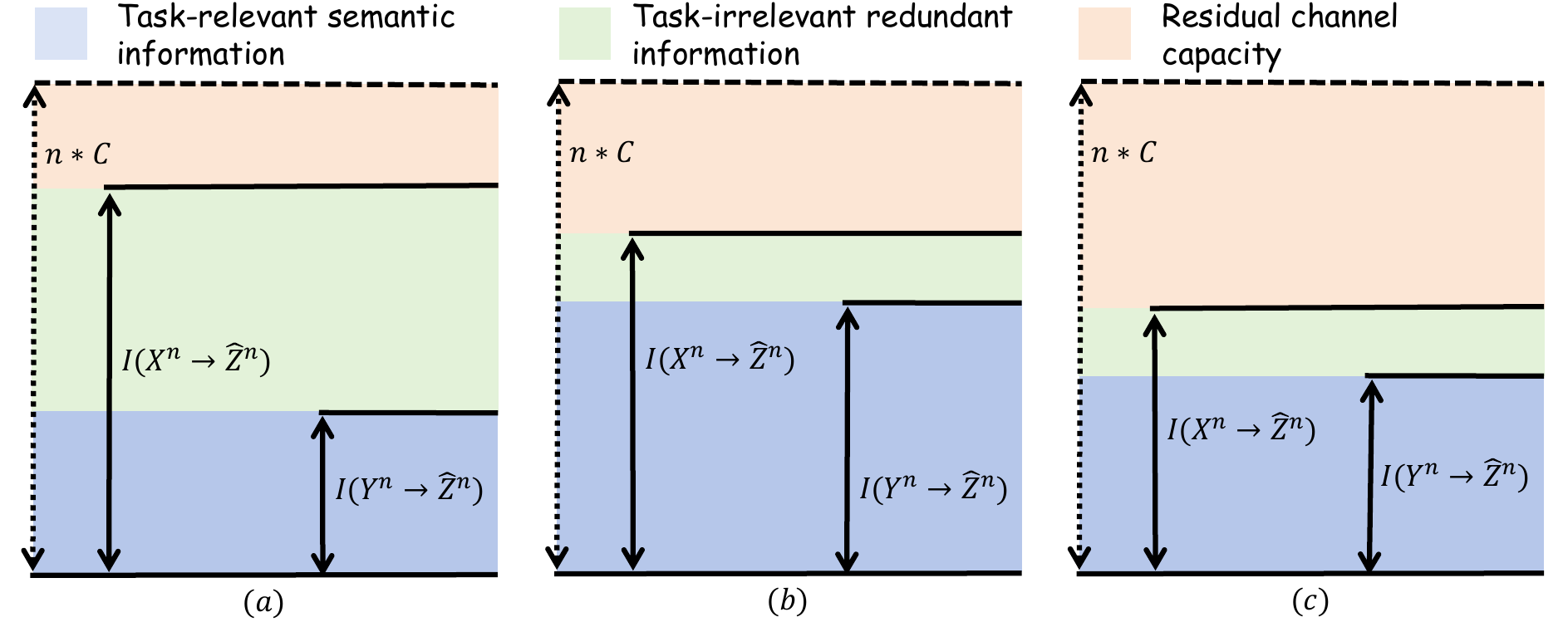}
    \caption{(a) Illustration of the data processing inequality in Eq. \eqref{eq:data_proc}, showing the relationship among the task-relevant semantic information $I(Y^n \rightarrow \widehat{Z}^n)$, the task-irrelevant redundant information $I(X^n \rightarrow \widehat{Z}^n) - I(Y^n \rightarrow \widehat{Z}^n)$, and the channel capacity not utilized for conveying information about the source information $X^n$, given by $n \cdot C - I(X^n \rightarrow \widehat{Z}^n)$; (b) The DIB with a smaller value of $\beta$ emphasizes semantic preservation, resulting in more semantic information and a higher compression cost; (c) The DIB with a larger value of $\beta$ emphasizes compression, leading to less semantic information and lower compression cost.}
    \label{fig:optimization}
\end{figure}

The proposed DIB framework, i.e., problem \eqref{eq:DIB}, inherently satisfies the data processing inequality established in Theorem \ref{them:data_inequality}. In particular, the parameter $\beta$ controls the trade-off between semantic relevance and compression cost, with the objective of retaining and transmitting task-relevant incremental semantic information while eliminating task-irrelevant redundancy.
The impact of $\beta$ is illustrated in Fig. \ref{fig:optimization}, and further discussed in Remark \ref{rem:beta_DIB}.

\begin{myRem}[How to Select the Parameter $\beta$ in the DIB Framework?]\label{rem:beta_DIB} To gain further insight into the role of $\beta$, we decompose the channel capacity $n\cdot C$ ($n$ corresponds to timestamp $t_n$) into task-relevant semantic information, task-irrelevant redundancy information, and residual channel capacity, as follows  
\begin{equation}
    \begin{aligned}
    &n \cdot C \\
    =&\underbrace{I(Y^n \rightarrow \widehat{Z}^n)}_{\text{Semantic Relevance}} + \underbrace{I(X^n \rightarrow \widehat{Z}^n) - I(Y^n \rightarrow \widehat{Z}^n)}_{\text{Redundant Information}} \\
    &+ \underbrace{n\cdot C - I(X^n \rightarrow \widehat{Z}^n)}_{\text{Residual Channel Capacity}}.
    \end{aligned}
\end{equation}
Particularly, the term $I(Y^n \rightarrow \widehat{Z}^n)$ quantifies the amount of information contained in the semantic encodings $\widehat{Z}^n$ that is relevant to the targets $Y^n$, the term $I(X^n \rightarrow \widehat{Z}^n) - I(Y^n \rightarrow \widehat{Z}^n)$ measures the amount of transmitted information that is irrelevant to $Y^n$, and the term $n\cdot C - I(X^n \rightarrow \widehat{Z}^n)$ denotes the channel capacity not utilized for transmitting information about the source information $X^n$, with their relationships illustrated in Fig. \ref{fig:optimization}(a).
Moreover, the problem \eqref{eq:DIB} indicates that a smaller value of $\beta$ places greater emphasis on semantic relevance (i.e., $I(Y^n \rightarrow \widehat{Z}^n)$), whereas a larger value of $\beta$ assigns a higher penalty to the compression cost (i.e., $I(X^n \rightarrow \widehat{Z}^n)$) and therefore favors compression, as illustrated in Fig. \ref{fig:optimization}(b) and Fig. \ref{fig:optimization}(c).
Note that regardless of the value of $\beta$, maximizing problem \eqref{eq:DIB} encourages the retention of more task-relevant semantic information and the removal of more task-irrelevant redundancy.
\end{myRem}

\section{Variational Directed Information Bottleneck for Task-Oriented Communication}\label{se:VDIBforSe}
In this section, we develop a variational formulation of the proposed DIB optimization problem and derive its variational upper bound, which serves as the theoretical foundation for the design of task-oriented communication systems.
According to Definition~\ref{def:directed_information}, the DIB problem \eqref{eq:DIB} can be equivalently reformulated as
\begin{align}
    &\max_{P(\widehat{z}^n \| x^n)} I(Y^n \rightarrow \widehat{Z}^n) - \beta I(X^n \rightarrow \widehat{Z}^n)  \\
    =&-\min_{P(\widehat{z}^n \| x^n)} \sum_{k=1}^n \left[ I(Y^k; \widehat{Z}_k | \widehat{Z}^{k-1}) - \beta  I(X^k; \widehat{Z}_k | \widehat{Z}^{k-1}) \right] \\
    =&\min_{P(\widehat{z}^n \| x^n)} \sum_{k=1}^n \left[ -H(\widehat{Z}_k | \widehat{Z}^{k-1}) + H(\widehat{Z}_k| Y^k, \widehat{Z}^{k-1}) \right] \nonumber \\
    &+ \beta \sum_{k=1}^n \left[ H(\widehat{Z}_k | \widehat{Z}^{k-1}) - H(\widehat{Z}_k | X^k, \widehat{Z}^{k-1}) \right]  \\
    =&\min_{P(\widehat{z}^n \| x^n)} \sum_{k=1}^n \big[(\beta - 1) H(\widehat{Z}_k | \widehat{Z}^{k-1}) \\
    &+ H(\widehat{Z}_k | Y^k, \widehat{Z}^{k-1}) - \beta H(\widehat{Z}_k | X^k, \widehat{Z}^{k-1}) \big],
\end{align} 
where the involved conditional entropy terms are given by
\begin{align}
    H(\widehat{Z}_k | \widehat{Z}^{k-1}) &= -\mathbb{E}_{P(\widehat{z}^k)} [\log P(\widehat{z}_k | \widehat{z}^{k-1})], \label{eq:entropy_first} \\
    H(\widehat{Z}_k | Y^k, \widehat{Z}^{k-1}) &= - \mathbb{E}_{P(y^k, \widehat{z}^k)}[\log P(\widehat{z}_k | y^k, \widehat{z}^{k-1})], \label{eq:entropy_second} \\
    H(\widehat{Z}_k | X^k, \widehat{Z}^{k-1}) &= - \mathbb{E}_{P(x^k, \widehat{z}^k)} [\log P(\widehat{z}_k | x^k, \widehat{z}^{k-1})]. \label{eq:entropy_third} 
\end{align}

Although the above reformulation explicitly characterizes the DIB framework in terms of conditional entropies, the resulting entropy terms are generally intractable since the underlying conditional distributions are unavailable. 
To obtain a tractable objective, we introduce a set of variational distributions to approximate these conditional distributions and derive a variational upper bound.
The following theorem establishes this bound.
\begin{myThe}\label{them:bound_DIB}The directed information bottleneck (DIB) objective is given by
\begin{equation}
    \mathcal{L}_{\mathrm{DIB}} = -I(Y^n \rightarrow \widehat{Z}^n) + \beta I(X^n \rightarrow \widehat{Z}^n).
\end{equation}
Let $\mathcal{A}_{\theta}(\widehat z_k|x^k,\widehat z^{k-1})$, $\mathcal{B}_{\phi}(y_k|\widehat z^k)$, and $r_{\psi}(\widehat z_k|\widehat z^{k-1})$ denote the causal semantic encoder, the variational decoder, and the autoregressive prior, respectively.
Then, the DIB objective is upper-bounded by the following variational objective
\begin{equation}\label{eq:VDIB}
\begin{aligned}
    \mathcal{L}_{\mathrm{VDIB}} &= \sum_{k=1}^n \underbrace{\mathbb{E}_{P(x^k,y^k)} \{ \mathbb{E}_{\mathcal{A}_{\theta}(\widehat{z}^k \| x^k)} [ -\log \mathcal{B}_{\phi}(y_k | \widehat{z}^k) ] \}}_{\text{Task Loss}} \\
    &+ \beta \sum_{k=1}^n \underbrace{D_{\mathrm{KL}} (\mathcal{A}_{\theta}(\widehat{z}_k | x^k,\widehat{z}^{k-1})\| r_{\psi}(\widehat{z}_k \mid \widehat{z}^{k-1}) )}_{\text{Information Regularization}},
\end{aligned}
\end{equation}
where $D_{\mathrm{KL}}(* \| *)$ denotes the Kullback--Leibler (KL) divergence, the causally conditioned distribution $\mathcal{A}_{\theta}(\widehat{z}^k \| x^k)$ is given by
\begin{equation}
    \mathcal{A}_{\theta}(\widehat{z}^k \| x^k) = \prod_{t=1}^k \mathcal{A}_{\theta}(\widehat{z}_t | x^t, \widehat{z}^{t-1}), 
\end{equation}
and the received semantic encoding over the AWGN channel is expressed as $\widehat{z}_k = z_k + \varepsilon_k,~\varepsilon_k \sim \mathcal{N}(\bm{0}, \sigma^2 \bm{I})$. 
Particularly, the resulting upper bound admits a rate-distortion decomposition, where the task loss acts as the distortion term and the KL regularizer quantifies the information rate, thus capturing the trade-off between task relevance and causal representation compression.
\end{myThe}

\begin{myProo} 
The proof is given in Appendix \ref{app:theorem_2}.
\end{myProo}

According to the Theorem \ref{them:bound_DIB}, we consider the variational optimization problem, i.e., $\min_{\theta, \phi, \psi} \mathcal{L}_{\mathrm{VDIB}}$, instead of directly solving $\min_{P(\widehat{z}^n \| x^n)} \mathcal{L}_{\mathrm{DIB}}$.
Since $\mathcal{L}_{\mathrm{VDIB}}$ serves as an upper bound of $\mathcal{L}_{\mathrm{DIB}}$ (i.e., $\mathcal{L}_{\mathrm{DIB}} \leq \mathcal{L}_{\mathrm{VDIB}}$), minimizing the variational objective provides a tractable surrogate to provide an approximate solution to the original DIB objective, i.e.,
\begin{equation}
    \min_{P(\widehat{z}^n \| x^n)} \mathcal{L}_{\mathrm{DIB}} \approx \min_{\theta, \phi, \psi} \mathcal{L}_{\mathrm{VDIB}}.
\end{equation}

Furthermore, we discuss how to compute the upper bound $\mathcal{L}_{\mathrm{VDIB}}$ in practice. The expectations (i.e., $\mathbb{E}[\ast]$) involved in the proposed variational objective can be efficiently approximated via Monte Carlo sampling, as their exact evaluation over continuous random variables generally involves intractable integrals.
Specifically, for the task loss term in $\mathcal{L}_{\mathrm{VDIB}}$, we sample a mini-batch of $N$ temporally correlated data-target sequences from $P(x^k, y^k)$, denoted by
\begin{equation}
    \{ (x^{k,(i)}, y^{k,(i)})\}_{i=1}^N, \quad \forall (x^{k,(i)},y^{k,(i)}) \sim P(x^k,y^k).
\end{equation}
Accordingly, the outer expectation with respect to $P(x^k,y^k)$ (i.e., $\mathbb{E}_{P(x^k, y^k)}[\ast]$) is approximated by the empirical average as
\begin{equation}
\begin{aligned}
    &\mathbb{E}_{P(x^k,y^k)}\{\mathbb{E}_{\mathcal{A}_{\theta}(\widehat{z}^k \| x^k)} [-\log \mathcal{B}_{\phi}(y_k | \widehat{z}^k)] \} \\
    \simeq & \frac{1}{N} \sum_{i=1}^N \mathbb{E}_{\mathcal{A}_{\theta}(\widehat{z}^k \| x^{k,(i)})} [-\log \mathcal{B}_{\phi}(y_k^{(i)} | \widehat{z}^k)].
\end{aligned}
\end{equation}
Next, for each sampled data-target sequence, we draw $M$ semantic encoding sequences $\{ z^{k,(i,j)} \}_{j=1}^M$ from the encoder $\mathcal{A}_{\theta}(z^k \| x^{k,(i)})$, 
i.e., $\forall z^{k,(i,j)} \sim \mathcal{A}_{\theta}(\widehat{z}^k \| x^{k,(i)})$.
Each sampled semantic encoding is then transmitted over the physical channel, yielding the received semantic encodings $\widehat{z}^{k,(i,j)} =  z^{k,(i,j)} + \varepsilon_k$.
Thereby, the inner expectation with respect to the semantic encoder (i.e., $\mathbb{E}_{\mathcal{A}_{\theta}(\widehat{z}^k \| x^{k,(i)})}[\ast]$) can be approximated utilizing Monte Carlo sampling as 
\begin{equation}
\begin{aligned}
    &\mathbb{E}_{P(x^k,y^k)}\{\mathbb{E}_{\mathcal{A}_{\theta}(\widehat{z}^k \| x^k)} [-\log \mathcal{B}_{\phi}(y_k | \widehat{z}^k)] \} \\
    \simeq & \frac{1}{N} \frac{1}{M} \sum_{i=1}^N \sum_{j=1}^M -\log \mathcal{B}_{\phi}(y_k^{(i,j)}| \widehat{z}^{k,(i,j)}).
\end{aligned}
\end{equation}
Similarly, the information regularization term in Eq. \eqref{eq:VDIB} (i.e., KL-divergence $D_{\mathrm{KL}}(\ast \| \ast)$) does not have a closed-form expression, and is therefore estimated via Monte Carlo sampling as follows
\begin{equation}
\begin{aligned}
    &D_{\mathrm{KL}}(\mathcal{A}_{\theta}(\widehat{z}_k | x^k, \widehat{z}^{k-1}) \| r_{\psi}(\widehat{z}_k | \widehat{z}^{k-1})) \\
    \simeq & \frac{1}{N} \frac{1}{M} \sum_{i=1}^N \sum_{j=1}^M  \log \frac{\mathcal{A}_{\theta}(\widehat{z}_k^{(i,j)} | x^{k,(i)}, \widehat{z}^{k-1, (i,j)})}{r_{\psi}(\widehat{z}_k^{(i,j)} | \widehat{z}^{k-1, (i,j)})}.
\end{aligned}
\end{equation}
Combining the above approximations, we obtain the following Monte Carlo estimator of $\mathcal{L}_{\mathrm{VDIB}}$, 
\begin{equation}\label{eq:Monte_Carlo_sampling}
    \begin{aligned}
        \mathcal{L}_{\mathrm{VDIB}} \simeq& \frac{1}{N} \frac{1}{M} \sum_{k=1}^n \sum_{i=1}^N \sum_{j=1}^M \Big\{ -\log \mathcal{B}_{\phi}(y_k^{(i)} | \widehat{z}^k) \\
        &+ \log \frac{\mathcal{A}_{\theta}(\widehat{z}_k^{(i,j)} | x^{k,(i)}, \widehat{z}^{k-1, (i,j)})}{r_{\psi}(\widehat{z}_k^{(i,j)} | \widehat{z}^{k-1, (i,j)})} \Big\}.
    \end{aligned}
\end{equation}

\begin{figure*}[!h]
    \centering
    \includegraphics[width=0.98\textwidth]{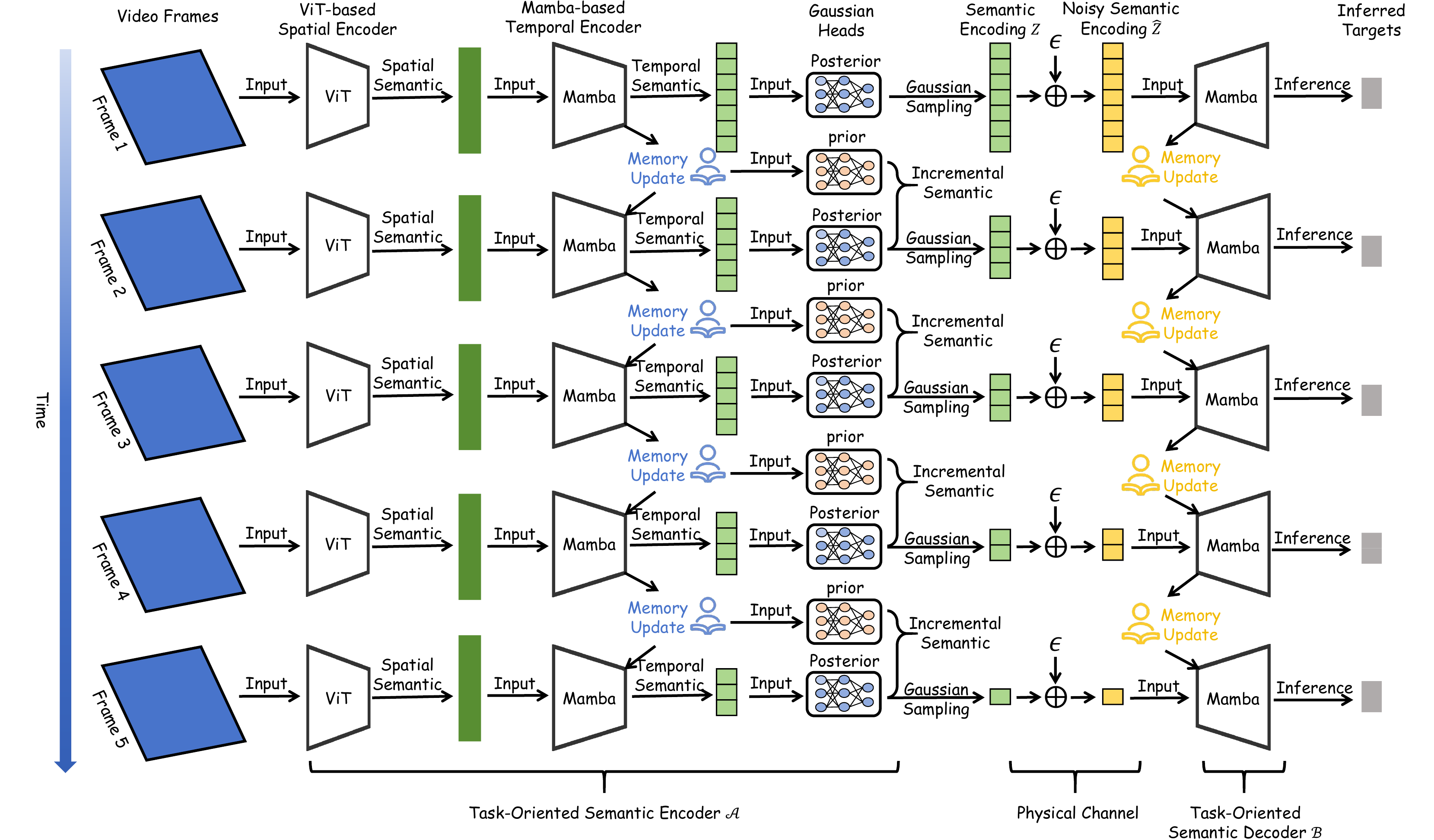}
    \caption{The architecture of the proposed task-oriented communication system, VDIB-JSCC, in which the task-oriented encoder $\mathcal{A}_{\theta}$ and decoder $\mathcal{B}_{\phi}$ are jointly optimized via end-end training. At the transmitter side, the encoder first extracts the spatial semantics of each incoming video frame using a ViT, then employs a Mamba-based temporal encoder to capture causal temporal dependencies and maintain a shared semantic memory, and finally parameterizes a variational posterior distribution to encode only the incremental task-relevant semantics into a compact latent representation for wireless transmission. Upon receiving the noisy semantic embedding vectors, the receiver employs a Mamba-based decoder to perform downstream inference tasks while maintaining receiver-side semantic memory through its latent state.}
    \label{fig:AI_model}
\end{figure*}

\section{Variational Directed Information Bottleneck for Task-Oriented Incremental Semantic Coding}\label{sec:VDIT_ISC}
In this section, we present the proposed variational directed information bottleneck joint source-channel coding (VDIB-JSCC) framework for task-oriented communication, as illustrated in Fig. \ref{fig:AI_model}.
Unlike conventional variational autoencoders~\cite{kingma2022autoencodingvariationalbayes,alemi2019deepvariationalinformationbottleneck,9606667,10845817} that employ a fixed Gaussian prior, VDIB-JSCC introduces a memory-conditioned autoregressive prior, whose parameters are dynamically determined by the historical semantic memory maintained by the Mamba module. 
Consequently, the prior is dynamically adapted to the evolving semantic context of the video stream rather than remaining static throughout the encoding process.
Specifically, a Vision Transformer (ViT)~\cite{dosovitskiy2021imageworth16x16words,10.1145/3505244} first extracts frame-wise spatial semantic representations, whose temporal dependencies are captured by a Mamba-based causal state-space model~\cite{10.1007/978-3-031-73347-5_14}. 
Based on the resulting temporal semantic, a variational posterior encoder infers the task-oriented latent representation from the current observation and historical semantic information, while a dynamic prior network predicts the latent distribution solely from previously transmitted semantic representations. Accordingly, the information regularization term in $\mathcal{L}_{\mathrm{VDIB}}$ (i.e., Eq. (\ref{eq:VDIB})) measures the discrepancy between the posterior and the memory-conditioned prior, thereby quantifying the incremental semantic information introduced by the current video frame.
This hierarchical design progressively removes task-irrelevant redundancy information in both spatial and temporal dimensions, leading to more efficient semantic transmission for temporally correlated video streams.
Next, we briefly review the fundamentals of ViT and Mamba, followed by a detailed description of the proposed VDIB-JSCC framework.
\subsection{Background on Vision Transformer and Mamba}
Vision Transformer (ViT) is a Transformer-based architecture designed to learn spatial semantic representations from visual inputs~\cite{dosovitskiy2021imageworth16x16words,10.1145/3505244}. 
Specifically, it partitions an input image into a sequence of patches and processes them as tokens through the self-attention mechanism, thereby capturing long-range spatial dependencies to extract task-relevant spatial semantic information while suppressing redundant visual details.
Since each video frame can be regarded as an independent image, ViT naturally provides discriminative frame-wise semantic representations for downstream task-oriented communication.

Although ViT can also be extended to spatiotemporal modeling by jointly attending to tokens from multiple frames, its quadratic computational complexity with respect to the sequence length makes it less suitable for long streaming sequences~\cite{neimark2021video,NEURIPS2022_416f9cb3,feichtenhofer2019slowfast}. Moreover, the global attention mechanism does not explicitly preserve the evolution of causal states, which is required to construct the autoregressive prior in the proposed VDIB framework.
Mamba~\cite{10.1007/978-3-031-73347-5_14} addresses these limitations by employing a selective state-space model, where the hidden state (referred to as memory in our setting) is  incrementally updated according to newly arriving inputs. 
This causal state-space formulation enables efficient modeling of long-range temporal dependencies with linear computational complexity, while naturally supporting incremental semantic extraction in streaming scenarios.
Consequently, Mamba is used here to serve as an effective temporal backbone for parameterizing the autoregressive prior $r_{\psi}$, where the causal state evolution captures temporal dependencies among spatial semantic representations and facilitates directed semantic information modeling over time.

\subsection{VDIB-JSCC}
We propose a novel variational directed information bottleneck joint source-channel coding (VDIB-JSCC) framework that extracts task-relevant incremental semantics from streaming video data and transmits only these incremental semantic representations, thereby substantially reducing the communication load.
As shown in Fig. \ref{fig:AI_model}, the proposed VDIB-JSCC framework is optimized through end-to-end joint training of the transmitter-side encoder $\mathcal{A}_{\theta}$ and the receiver-side decoder $\mathcal{B}_{\phi}$.
Specifically, $\mathcal{A}_{\theta}$ consists of a ViT-based spatial encoder, a Mamba-based temporal encoder, and two Gaussian heads, where one Gaussian head parameterizes the variational posterior distribution and the other parameterizes the dynamic prior distribution. The receiver-side decoder $\mathcal{B}{\phi}$ employs a Mamba-based semantic decoder for task-oriented information reconstruction.
In particular, ViT extracts task-relevant spatial semantic information from individual video frames, while Mamba captures temporal dependencies across frames and maintains the semantic memory. 
Based on the semantic memory and temporal representations generated by Mamba, the posterior and prior Gaussian heads parameterize the causal semantic encoder and autoregressive prior distributions, respectively.
The resulting KL divergence in the information regularization term of $\mathcal{L}_{\mathrm{VDIB}}$ characterizes the incremental semantic information by measuring the discrepancy between the posterior and prior distributions.

Specifically, at the transmitter side, we develop a task-oriented encoder $\mathcal{A}_{\theta}$ to extract incremental semantic information from temporally evolving video streams, while removing task-irrelevant redundancy with respect to the historical transmitter-side knowledge base (e.g., $D_{k-1}^t$ up to timestamp $t_k$).
Instead of explicitly maintaining a large-scale knowledge base, we exploit the latent states of Mamba as an implicit semantic memory to compactly preserve historical information.
To be more specific, for each incoming video frame $x_k$, the ViT-based spatial encoder first extracts the task-oriented spatial semantic information, given by
\begin{equation}\label{eq:spatial_semantic}
    s_k = \mathrm{ViT}_{\theta_v}(x_k), ~k = 1, 2, \cdots, n. 
\end{equation}
Subsequently, the extracted spatial semantic information $s_k$ is processed by the Mamba-based temporal encoder to capture temporal dependencies across consecutive frames and update the semantic memory $m^t_k$.
Based on the updated semantic memory, the temporal encoder further extracts the incremental semantic representation $\Delta s_k$ associated with the current frame.
The semantic memory update and incremental semantic extraction are formulated as
\begin{align}
    m_k^t &= A_t m_{k-1}^t + B_t s_k, ~ k = 1,2, \cdots, n, \label{eq:mamba_memory_updata} \\
    \Delta s_k &= C_t m_k^t,~ k =1, 2, \cdots, n, \label{eq:mamba_temporal_sematic_updata}
\end{align}
where $A_t$, $B_t$, and $C_t$ denote the learnable state-space parameters. For simplicity, the above temporal modeling process is denoted as
\begin{equation}
    (m_k^t, \Delta s_k) = \mathrm{Mamba}_{\theta_m}(m^t_{k-1}, s_k), ~ k=1, 2, \cdots, n,
\end{equation}
where $\theta_m = \{A_t,B_t,C_t \}$ represents the trainable parameters of the Mamba-based temporal encoder.
To obtain the semantic representation required by $\mathcal{L}_{\mathrm{VDIB}}$ (discussed in Section \ref{se:VDIBforSe}), a Gaussian head with parameters $\theta_g$ is employed to parameterize the variational posterior distribution by estimating the mean and variance from the extracted incremental semantic representation $\Delta s_k$. Specifically, the posterior distribution is parameterized as
\begin{equation}\label{eq:parameter_Gaussian_encoder}
    (\mu_k^q, \sigma_k^q)=\mathrm{Gaussian}_{\theta_g}(\Delta s_k),~ k=1,2, \cdots, n.
\end{equation}
Following the reparameterization trick~\cite{alemi2019deepvariationalinformationbottleneck,diederik2014auto}, the transmitted semantic encoding $z_k$ is sampled from the posterior distribution as
\begin{equation}\label{eq:reparameterization_trick_z_k}
    z_k = \mu_k^q + \sigma_k^q \cdot \epsilon_k, ~ k =1, 2, \cdots, n, 
\end{equation}
where $\epsilon_k \sim \mathcal{N}(\bm{0}, \bm{I})$ denotes the Gaussian noise, which enables differentiable sampling of the stochastic semantic representation and allows efficient optimization of the information regularization term associated with the posterior-prior discrepancy in $\mathcal{L}_{\mathrm{VDIB}}$. 
Therefore, the semantic encoder $\mathcal{A}_{\theta}$ consists of the ViT-based spatial encoder, the Mamba-based temporal encoder, and the posterior Gaussian head, with trainable parameters $\theta = \{\theta_v, \theta_m, \theta_g \}$.
In addition, conditioned on the previous semantic memory $m_{k-1}^t$, a separate Gaussian head with parameters $\psi$ is introduced to parameterize the autoregressive prior distribution $r_{\psi}$, which is formulated as
\begin{equation}\label{eq:parameter_Gaussian_prior}
    (\mu_k^p, \sigma_k^p) = \mathrm{Gaussian}_{\psi}(m^t_{k-1}),~k=1,2,\cdots,n.
\end{equation}
Unlike a fixed prior distribution, the autoregressive prior predicts the latent semantic distribution conditioned on previously accumulated semantic knowledge, thereby enabling the model to capture newly emerged semantic information while avoiding redundant encoding of previously acquired semantics.

\begin{algorithm}[!t] 
    \caption{Training VDIB-JSCC} \label{alg:training_VDIT_JSCC}	
    \SetKwInOut{Input}{\textbf{Input}}
    \SetKwInOut{Output}{\textbf{Output}}
    \Input{$T$: Number of iterations; $N$: Batch size}
    \While{epoch $t=1$ to $T$}
    {
        Sample a mini-batch $\{(\bm{x}^{(i)},\bm{y}^{(i)})\}_{i=1}^{N}$, where $\bm{x}^{(i)}\triangleq x^{K,(i)}$ is a $K$-frame video sequence and $\bm{y}^{(i)} \triangleq y^{K,(i)} $ denotes the task labels.\\
        Initialize the transmitter- and receiver-side memories to $m_0^t = m_0^r = \mathbf{0}$. \\
        \While{$i = 1$ to $N$}
        {
            \While{$k=1$ to $K$}
            {
            Extract spatial semantic representation $s_k^{(i)}$ from $x_k^{(i)}$ via ViT-based spatial encoder, as defined in Eq. (\ref{eq:spatial_semantic}).  \\
            Update the transmitter-side semantic memory $m_k^{t,(i)}$ and obtain incremental semantic information $\Delta s_k^{(i)}$ via Mamba-based temporal encoder, as defined in Eqs. (\ref{eq:mamba_memory_updata}) and (\ref{eq:mamba_temporal_sematic_updata}). \\
            Parameterize the posterior distribution $\mathcal{A}(z_k^{(i)} | x^{k,(i)}, z^{k-1,(i)})$ based on $\Delta s_k^{(i)}$ via a Gaussian head, as defined in Eq. (\ref{eq:parameter_Gaussian_encoder}). \\
            Sample the incremental semantic encoding $z_k^{(i)}$ from the parameterized distribution $\mathcal{A}$ via the reparameterization trick, as defined in Eq. (\ref{eq:reparameterization_trick_z_k}). \\
            Parameterize the autoregressive prior distribution $r_{\psi}(z_k^{(i)} | z^{k-1,(i)})$ based on $m_{k-1}^{t,(i)}$ via a Gaussian head, as defined in Eq. (\ref{eq:parameter_Gaussian_prior}).    \\
            Transmit the semantic embedding $z_k^{(i)}$ over the physical channel with additive Gaussian noise  $\epsilon_k^{(i)} \sim \mathcal{N}(\bm{0}, \sigma^2 \bm{I})$, as defined in Eq. (\ref{eq:physical_channel}). \\
            Update the receiver-side semantic memory $m_k^{r,(i)}$ and predict the corresponding label $\widehat{y}_k^{(i)}$ via Mamba-based decoder, as defined in Eqs. (\ref{eq:mamba_decoder_mmemory}) and (\ref{eq:mamba_decoder}). \\
            }
        }
        Compute the loss $\mathcal{L}_{\mathrm{VDIB}}$ via Monte Carlo sampling, as defined in Eq. (\ref{eq:Monte_Carlo_sampling}).\\
        Update all trainable parameters $\{\theta, \psi, \phi \}$ via backpropagation. \\
    }
\end{algorithm}

At the receiver, a Mamba-based decoder is employed to perform the downstream task, where its latent state implicitly serves as the receiver-side semantic memory to retain previously received semantic knowledge (e.g., $D_{k-1}^r$ up to timestamp $t_k$).
Each semantic encoding $z_k$ is received through the AWGN channel, yielding the received noisy encoding $\widehat{z}_k$.
Given $\widehat{z}_k$, the decoding process is formulated as
\begin{equation}
    (m_k^r, \widehat{y}_k) = \mathrm{Mamba}_{\phi}(m_{k-1}^r, \widehat{z}_k),~k=1,2,\cdots, n,
\end{equation}
where the memory update and target inference are given by
\begin{align}
    m^r_k &= A_r m^r_{k-1} + B_r \widehat{z}_k, ~ k = 1,2, \cdots, n, \label{eq:mamba_decoder_mmemory} \\
    \widehat{y}_k &= C_r m_k^r, ~ k=1,2,\cdots, n, \label{eq:mamba_decoder}
\end{align}
where $\phi = \{ A_r, B_r, C_r\}$ denotes the trainable parameters of the receiver-side decoder, and $\widehat{y}_k$ is the inferred target at timestamp $t_k$.

The training details of the proposed VDIB-JSCC framework are summarized in Algorithm~\ref{alg:training_VDIT_JSCC}.

\section{Experimental Evaluation}\label{sec:experimental}
In this section, we evaluate the performance of VDIB-JSCC on two widely adopted image classification benchmarks, CIFAR-10 and CIFAR-100, and compare it with state-of-the-art task-oriented communication methods.
We further conduct a robustness study to evaluate the effects of key parameters, including the channel $\mathrm{PSNR}$, penalty parameter $\beta$, and communication latent dimension $d_z$.

\subsection{Experimental Setup}
All experiments are conducted using the same datasets and implementation environment. The simulations are implemented in Python and conducted on four NVIDIA RTX 5000 Ada GPUs, each with 32 GB of GPU memory.

\subsubsection{Datasets}
We evaluate VDIB-JSCC on two widely adopted image classification benchmarks, CIFAR-10 and CIFAR-100\footnote{\url{https://cave.cs.toronto.edu/kriz/cifar.html}}.
Both datasets contain 60,000 images of size $32\times32$, including 50,000 training images and 10,000 test images. CIFAR-10 consists of 10 object classes (e.g., airplane and bird), with 5,000 training and 1,000 test images per class. In contrast, CIFAR-100 contains 100 fine-grained classes (e.g., aquarium fish and flatfish), with 500 training and 100 test images per class, which poses a more challenging classification task due to the larger number of classes and greater inter-class similarity.
To construct streaming sequences, we randomly select $T$ images from the same class and arrange them sequentially. By this construction strategy, we create substantial task-relevant semantic redundancy among the $T$ images, providing a suitable setting for evaluating the effectiveness of the proposed task-oriented communication framework that transmits only incremental semantic information. For the state-of-the-art compared methods, each image is treated as an independent classification instance, and classification performance is evaluated at each time step as new images are received, ensuring a fair comparison with VDIB-JSCC.

\subsubsection{Compared Methods}
To the best of our knowledge, no existing task-oriented communication system further reduces communication load by transmitting only incremental semantic information.
We evaluate VDIB-JSCC against three state-of-the-art task-oriented communication methods, including deep joint source-channel coding (DeepJSCC)~\cite{8723589}, variable-length variational feature encoding (VL-VFE)~\cite{9606667}, and discrete task-oriented joint source-channel coding (DT-JSCC)~\cite{10159007}.
All considered methods adopt learning-based, end-to-end JSCC architectures, while VL-VFE and DT-JSCC are task-oriented communication systems based on the information bottleneck (IB) framework.
However, these methods treat the transmitted data samples as mutually independent, overlooking the correlations and previously acquired semantic information across consecutive samples. A brief description of the considered methods is provided below.
\begin{itemize}
    \item DeepJSCC~\cite{8723589} was originally developed for data-oriented communication systems, employing DNN-based encoders to directly map source data to channel input symbols. Its end-to-end learning architecture has also been adopted in task-oriented communication systems for efficient transmission of task-relevant semantic information. Following this paradigm, we train DeepJSCC using the cross-entropy loss.
    \item VL-VFE~\cite{9606667} is a learning-based task-oriented communication method that jointly optimizes feature extraction, source coding, and channel coding for edge inference. VL-VFE adopts the IB framework to formulate a rate-distortion tradeoff between the informativeness of encoded features and inference performance, while introducing a sparsity-inducing prior to prune inactive feature dimensions and reduce the number of transmitted symbols. Furthermore, VL-VFE employs a dynamic neural network to adaptively adjust the active feature dimensions according to channel conditions.
    \item DT-JSCC~\cite{10159007} is a task-oriented communication method that employs a robust information bottleneck (RIB) framework to balance the informativeness and robustness of encoded representations. This method encodes task-relevant semantic information into discrete representations and transmits them utilizing digital modulation, improving robustness to channel variations while maintaining compatibility with conventional digital communication systems.
\end{itemize}

\subsubsection{Evaluation Metrics}
Most existing task-oriented communication systems primarily focus on rate-distortion tradeoff~\cite{10159007,10845817,9606667,10554663,9955312}.
For example, in the IB framework given by Eq. (\ref{eq:IB}), the mutual information $I(Z;Y)$ measures the task-relevant information preserved in the encoding $Z$ and is closely related to task distortion, while $I(X;Z)$ quantifies the information rate of $Z$.
In our DIB framework, these terms are replaced by the directed information $I(Y^n \rightarrow \widehat{Z}^n)$ and $I(X^n \rightarrow \widehat{Z}^n)$, which quantify the task-relevant incremental information and the corresponding information rate provided by the noisy encoding $\widehat{Z}^n$, respectively.
Accordingly, our experiments evaluate the performance of task-oriented  communication in terms of task distortion (measured by task accuracy) and information rate (measured by compression cost).
Specifically, the time-step accuracy quantifies the classification performance achieved by task inference based on the information accumulated up to $t_k$ and is defined as 
\begin{equation}
    \mathrm{Acc}_{k} = \frac{1}{N} \sum_{i=1}^N  \mathbb{I}(\widehat{y}_{i,t_k} = y_i).
\end{equation}
Furthermore, the average time-step accuracy is formulated as $\mathrm{Acc} = \mathbb{E} [ \mathrm{Acc}_k ] =  \frac{1}{T} \sum_{k=1}^T \mathrm{Acc}_k $.

\subsubsection{Hyperparameter Setting}
For the proposed VDIB-JSCC, each input sequence consists of $T=8$ images, with each image resized to $(3,224,224)$. The images are first processed by $\mathrm{vit\_base\_patch16\_224}$\footnote{\url{https://huggingface.co/google/vit-base-patch16-224}}, which serves as the backbone of the spatial encoder to extract task-relevant spatial semantic information, with its pretrained parameters kept frozen throughout training.
The extracted spatial semantic information are subsequently processed by a Mamba module with a state-space dimension of $d_{\mathrm{state}}=16$, which captures temporal and causal dependencies to identify the incremental semantic information beyond previously acquired knowledge.
The resulting incremental semantic information is further encoded into a $d_z$-dimensional latent coding through Gaussian heads and transmitted over the physical channel.
Accordingly, the spatial semantic information, incremental semantic information, and communication latents have dimensions of $N\times T\times d_s$, $N\times T\times d_i$, and $N\times T\times d_z$, respectively, where $d_s=d_i=d_z=512$.
For the physical channel, the transmitted latent coding is passed through an additive white Gaussian noise (AWGN) channel, where Gaussian noise $\epsilon\sim\mathcal{N}(0,\sigma^2\mathbf{I})$ is added to the channel input coding. The channel PSNR is defined as
\begin{equation}
\mathrm{PSNR}=10\log_{10}\frac{P_{\max}}{\sigma^2}~(\mathrm{dB}),
\end{equation}
where $P_{\max}$ denotes the maximum power of the channel input coding.
Unless otherwise specified, the channel $\mathrm{PSNR}$ is preset to $\mathrm{PSNR} = 15$ dB during both training and testing.
Upon receiving the transmitted latent coding, the receiver feeds it into a Mamba-based decoder with $d_{\mathrm{state}} = 16$ to perform the classification task.
For the VDIB framework, we set the penalty parameter to $\beta=10^{-3}$ (see Remark~\ref{rem:beta_DIB}), which controls the trade-off between preserving task-relevant semantic information and compressing input information. The effects of $\mathrm{PSNR}$, $\beta$, and $d_z$ are further investigated in the subsequent experiments.

\begin{table}[t!]
	\centering
	\caption{Inference accuracy of VDIB-JSCC and three compared methods under different PSNR on CIFAR-10.}\label{tab:reslults_cifar-10}
	\begin{tabular}{c|cccccc}
		\toprule  
		\text{PSNR} & \text{3 dB} & \text{5 dB} & \text{10 dB} & \text{15 dB} & \text{20 dB} & \text{30 dB} \\
		\midrule
        DeepJSCC & 90.25 & 91.08 & 91.27 & 91.45 & 91.67 & 91.92\\
        VL-VFE & 90.81 &  91.24 & 91.52 & 91.79 & 91.90 & 92.01\\
        DT-JSCC & 91.68 & 91.84 & 91.91 & 91.93 & 92.14 & 92.15\\
        VDIB-JSCC & \textbf{98.87} & \textbf{99.04} & \textbf{99.08} & \textbf{99.09} & \textbf{99.09} & \textbf{99.11}\\
        \bottomrule 
	\end{tabular}
\end{table}

\begin{figure}[!t]
    \centering
    \subfloat[Task Accuracy on CIFAR-10]{
        \includegraphics[width=0.47\linewidth]{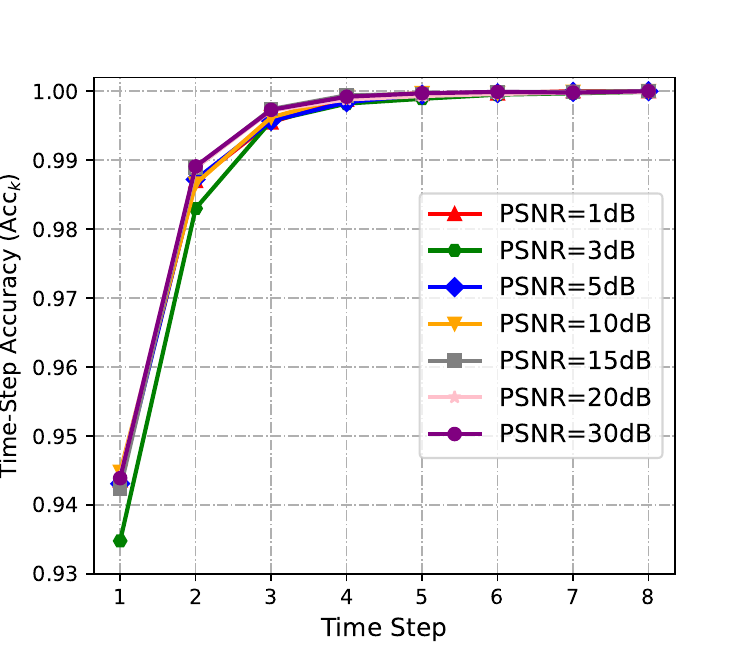}
        \label{fig:timeAcc_kl_a}
    }
    \hfill
    \subfloat[Compression Cost on CIFAR-10]{
        \includegraphics[width=0.47\linewidth]{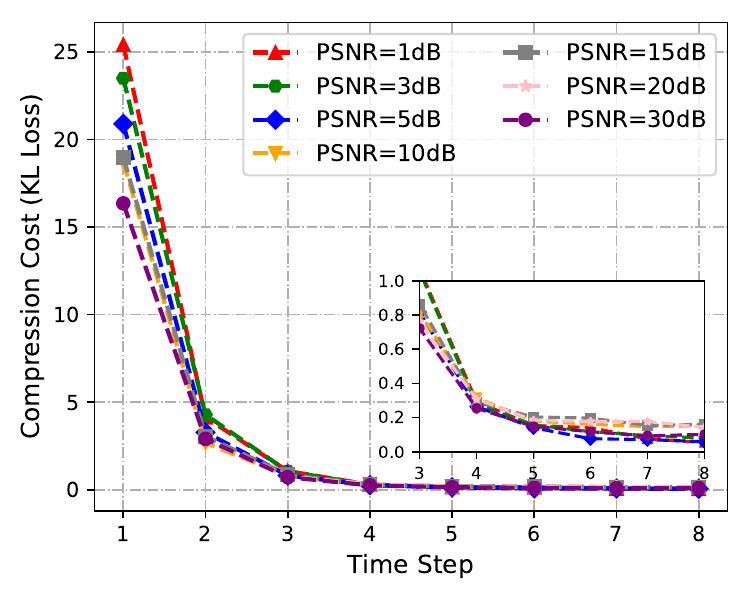}
        \label{fig:timeAcc_kl_b}
    }
    \hfill
    \subfloat[Task Accuracy on CIFAR-100]{
        \includegraphics[width=0.47\linewidth]{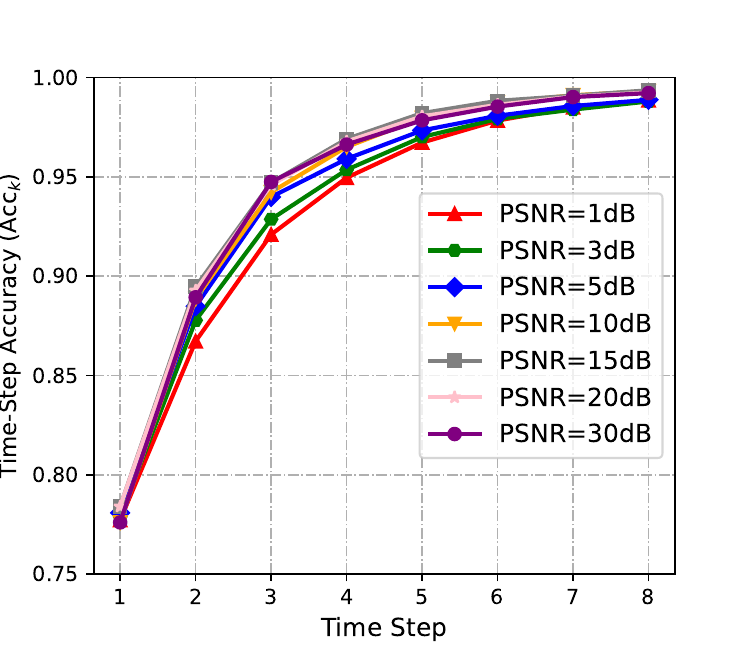}
        \label{fig:timeAcc_kl_c}
    }
    \hfill
    \subfloat[Compression Cost on CIFAR-100]{
        \includegraphics[width=0.47\linewidth]{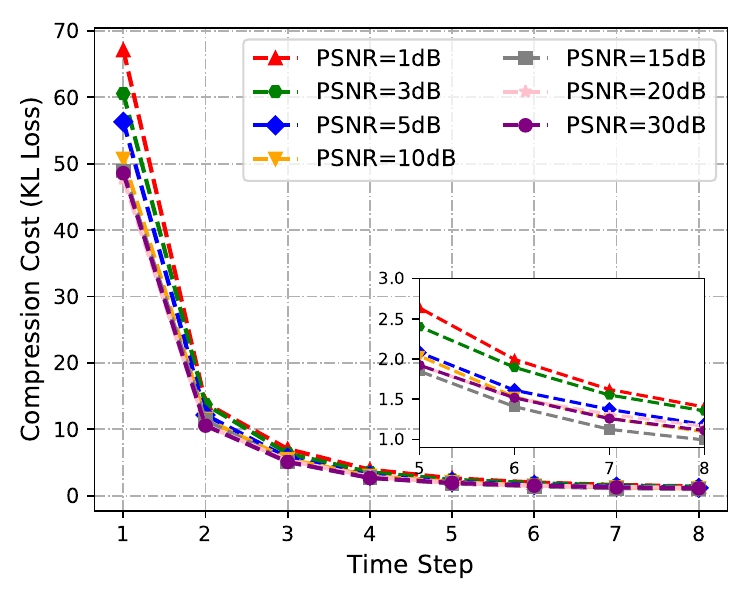}
        \label{fig:timeAcc_kl_d}
    }
    \caption{Performance of VDIB-JSCC under different $\mathrm{PSNR} \in \{1, 3, 5, 10, 15, 20, 30\}$ dB, where $\mathrm{PSNR}$ used for training and testing is the same for each curve, with $\beta = 10^{-3}$ and $d_z = 512$. (a) and (c) depict the top-1 task accuracy over time steps. (b) and (d) show the compression cost at each time step, i.e., the information regularization term in $\mathcal{L}_{\mathrm{VDIB}}$.}
    \label{fig:timeAcc_KLloss}
\end{figure}

\subsection{Main Results}
We evaluate the inference performance of VDIB-JSCC and the compared methods upon receiving the task-relevant semantic information transmitted through the physical channel, with the results reported in Tables~\ref{tab:reslults_cifar-10} and~\ref{tab:reslults_cifar-100}.
Across both CIFAR-10 and CIFAR-100, the proposed VDIB-JSCC consistently outperforms DeepJSCC, VL-VFE, and DT-JSCC in terms of task accuracy, achieving excellent task performance across different channel conditions.
For instance, at $\mathrm{PSNR} = 15$ dB, VDIB-JSCC achieves an average time-step accuracy of $99.09\%$ on CIFAR-10 and maintains a high accuracy of $94.24\%$ on CIFAR-100, demonstrating that effective task inference can be achieved by transmitting only incremental task-relevant information.
Moreover, at $\mathrm{PSNR} = 15$ dB, VDIB-JSCC achieves at least $7.16\%$ and $17.79\%$ gains over the compared methods on CIFAR-10 and CIFAR-100, respectively, in terms of average time-step accuracy.
This performance gain stems from its ability to capture temporal and causal correlations and exploit incrementally accumulated knowledge for task inference.

\begin{table}[t!]
	\centering
	\caption{Classification accuracy of VDIB-JSCC and three compared methods over different PSNR on CIFAR-100.}\label{tab:reslults_cifar-100}
	\begin{tabular}{c|cccccc}
		\toprule  
		\text{PSNR} & \text{3 dB} & \text{5 dB} & \text{10 dB} & \text{15 dB} & \text{20 dB} & \text{30 dB} \\
		\midrule
        DeepJSCC & 76.38 & 76.39 & 76.41 & 76.46 & 76.48 & 76.51\\
        VL-VFE & 76.41 & 76.43  & 76.47 & 76.49 & 76.52 & 76.58\\
        DT-JSCC & 76.50 & 76.52 & 76.55 & 76.57 & 76.61 & 76.62\\
        VDIB-JSCC & \textbf{93.24} & \textbf{93.66} & \textbf{94.03} & \textbf{94.36} & \textbf{94.24} & \textbf{94.06}\\
        \bottomrule 
	\end{tabular}
\end{table}

\begin{figure}[!t]
    \centering
    \subfloat[Avg. Task Accuracy on CIFAR-10]{
        \includegraphics[width=0.47\linewidth]{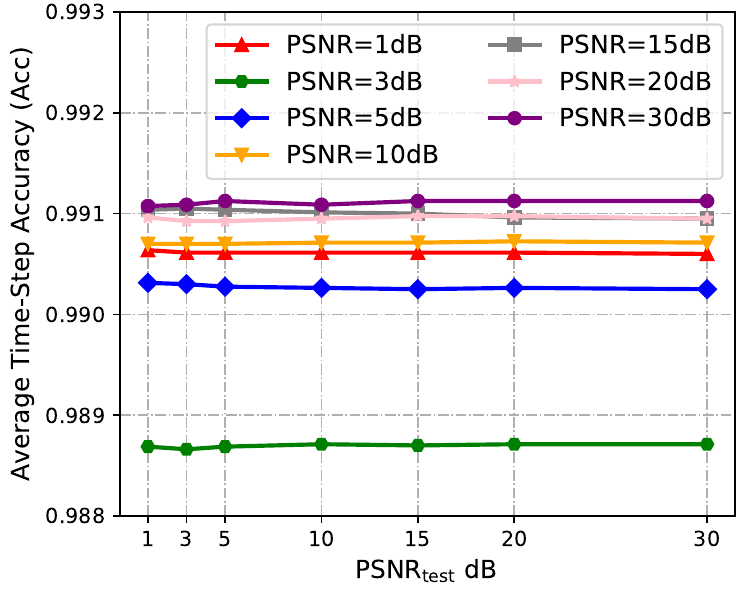}
        \label{fig:time_step_acc_a}
    }
    \hfill
    \subfloat[Avg. Task Accuracy on CIFAR-100]{
        \includegraphics[width=0.47\linewidth]{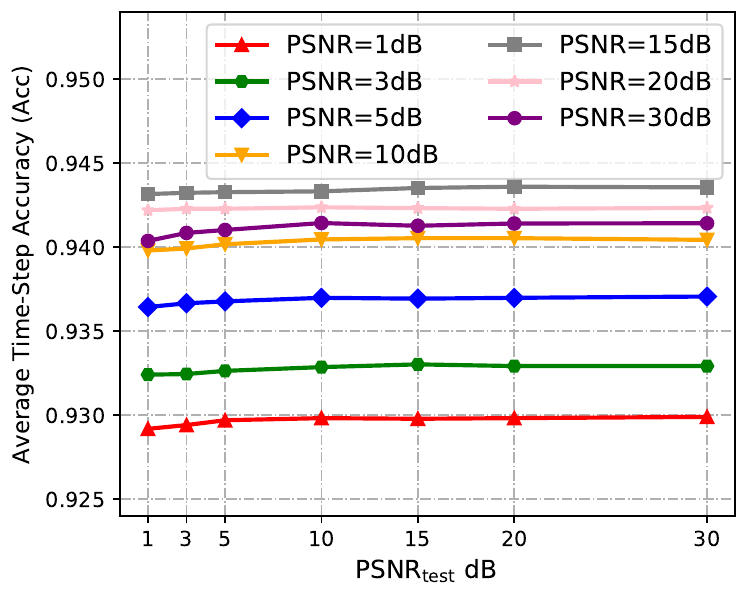}
        \label{fig:time_step_acc_b}
    }
    \caption{Performance of VDIB-JSCC under different AWGN channels with $\mathrm{PSNR} \in \{1, 3, 5, 10, 15, 20, 30\}$ dB, where each approach is trained at the same PSNR value with $\beta = 10^{-3}$ and $d_z = 512$.}
    \label{fig:time_step_acc}
\end{figure}

\begin{figure}[!t]
    \centering
    \subfloat[Task Accuracy on CIFAR-10]{
        \includegraphics[width=0.47\linewidth]{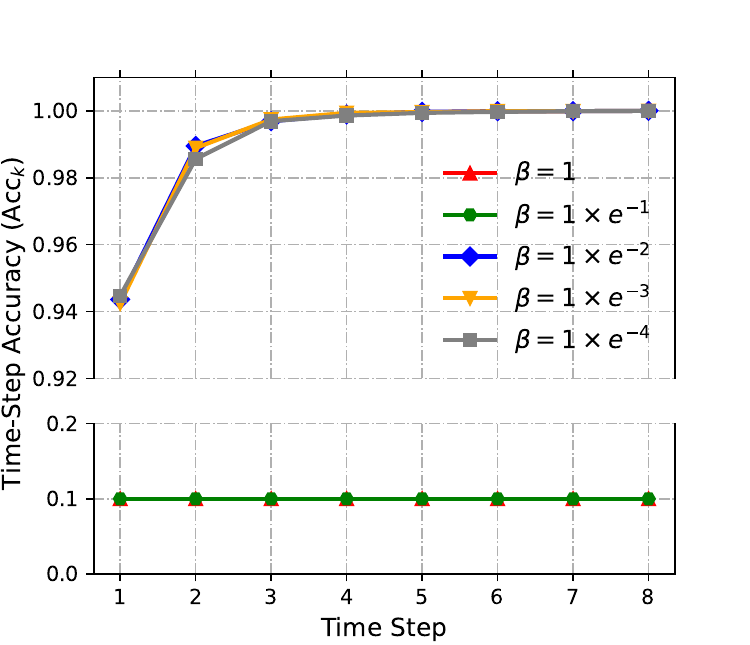}
        \label{fig:timeAcc_beta_kl_a}
    }
    \hfill
    \subfloat[Compression Cost on CIFAR-10]{
        \includegraphics[width=0.47\linewidth]{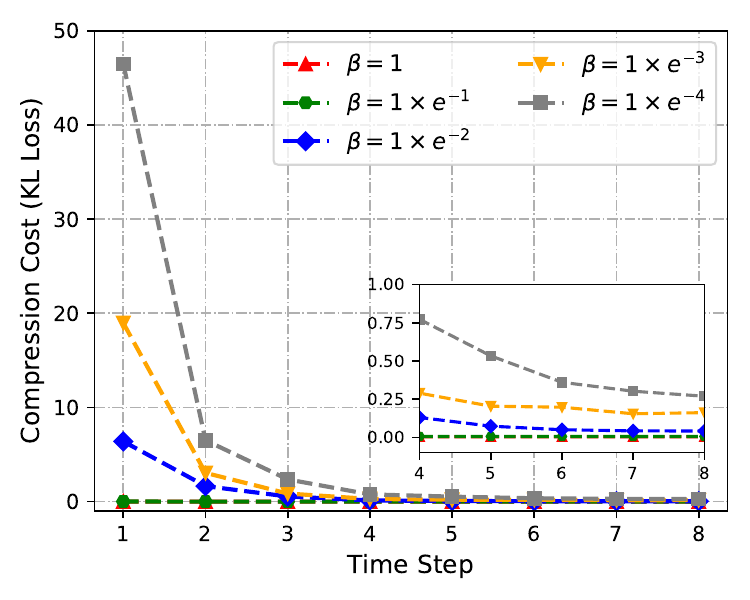}
        \label{fig:timeAcc_beta_kl_b}
    }
    \hfill
    \subfloat[Task Accuracy on CIFAR-100]{
        \includegraphics[width=0.47\linewidth]{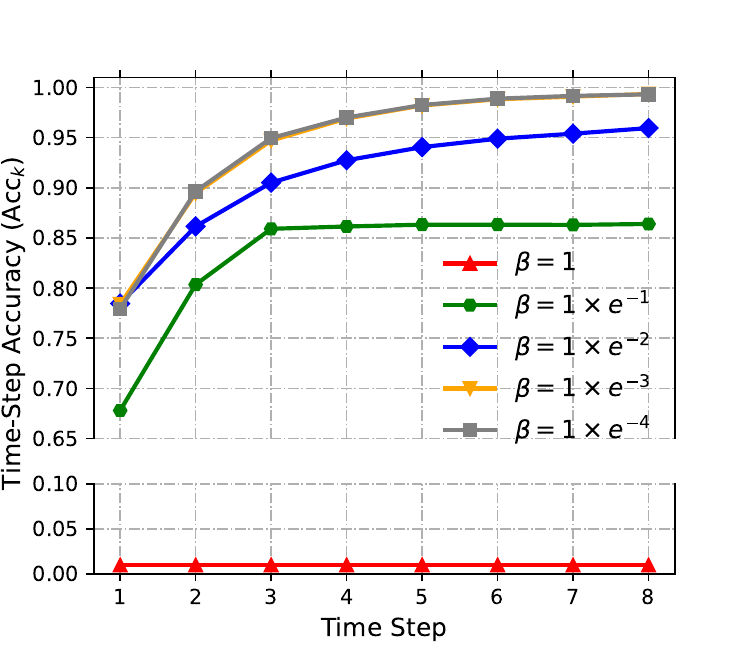}
        \label{fig:timeAcc_beta_kl_c}
    }
    \hfill
    \subfloat[Compression Cost on CIFAR-100]{
        \includegraphics[width=0.47\linewidth]{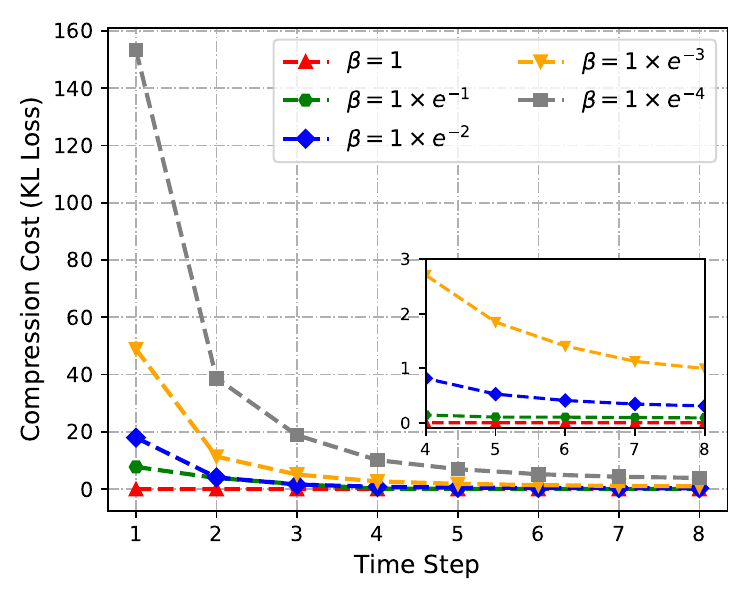}
        \label{fig:timeAcc_beta_kl_d}
    }
    \caption{Performance of VDIB-JSCC under different $\beta \in \{1, 10^{-1}, 10^{-2}, 10^{-3}, 10^{-4} \}$, where $\mathrm{PSNR} = 15$ dB and $d_z = 512$ are fixed for both training and testing. (a) and (c) depict the top-1 task accuracy at each time step as the received information accumulates, while (b) and (d) present the corresponding compression cost of the transmitted information.}
    \label{fig:timeAcc_beta_KLloss}
\end{figure}

\begin{figure}[!t]
    \centering
    \subfloat[ Compression Cost on CIFAR-10]{
        \includegraphics[width=0.47\linewidth]{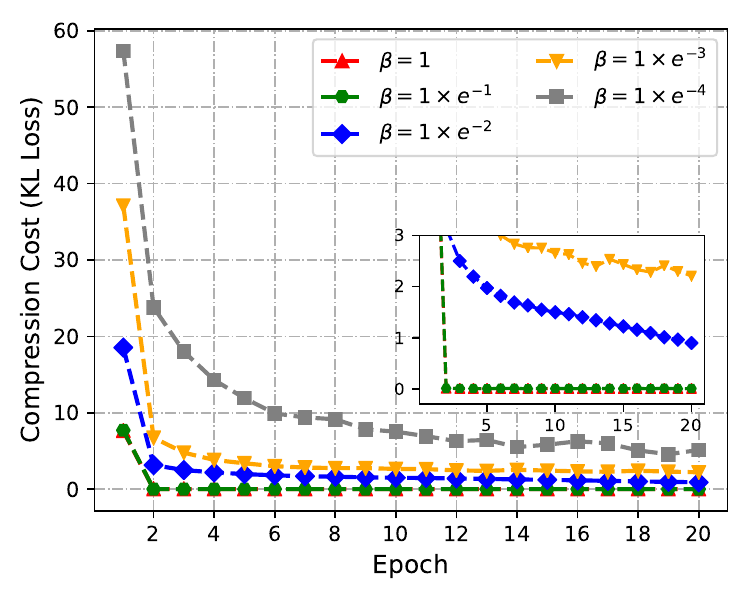}
        \label{fig:loss_beta_a}
    }
    \hfill
    \subfloat[Semantic Relevance on CIFAR-10]{
        \includegraphics[width=0.47\linewidth]{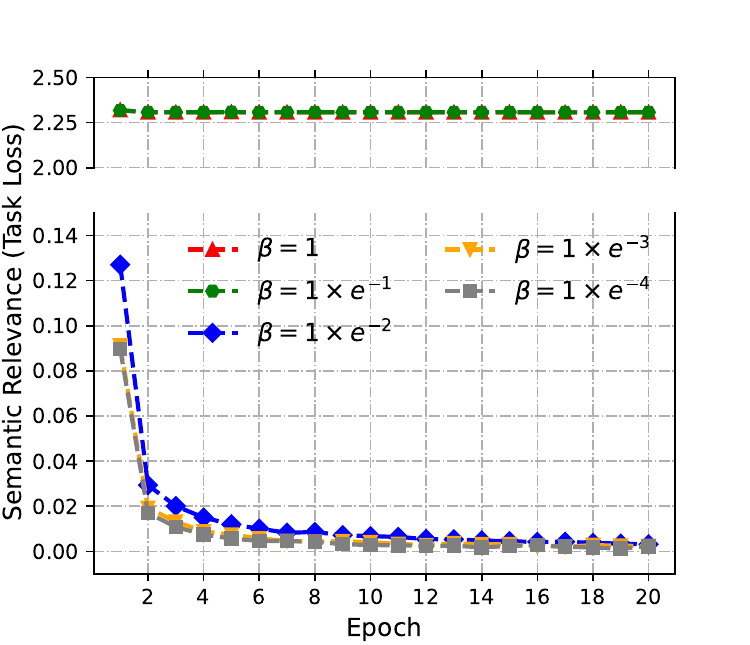}
        \label{fig:loss_beta_b}
    }
    \hfill
    \subfloat[Compression Cost on CIFAR-100]{
        \includegraphics[width=0.47\linewidth]{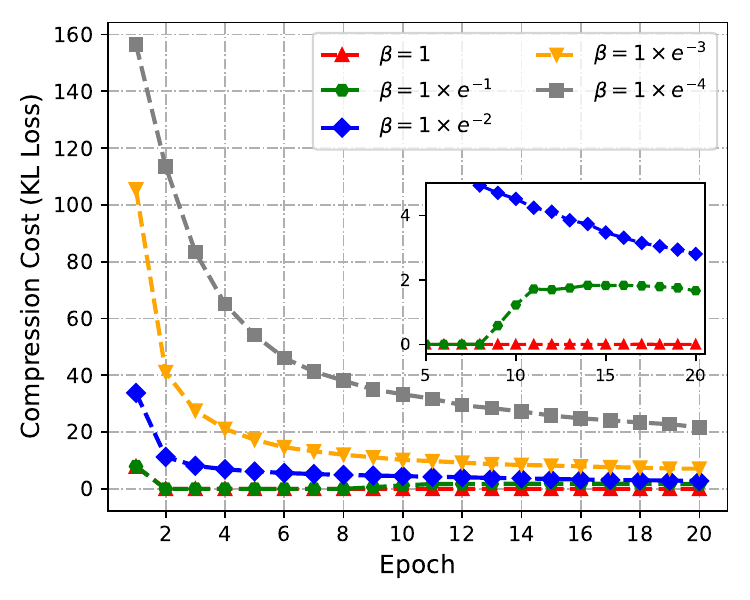}
        \label{fig:loss_beta_c}
    }
    \hfill
    \subfloat[Semantic Relevance on CIFAR-100]{
        \includegraphics[width=0.47\linewidth]{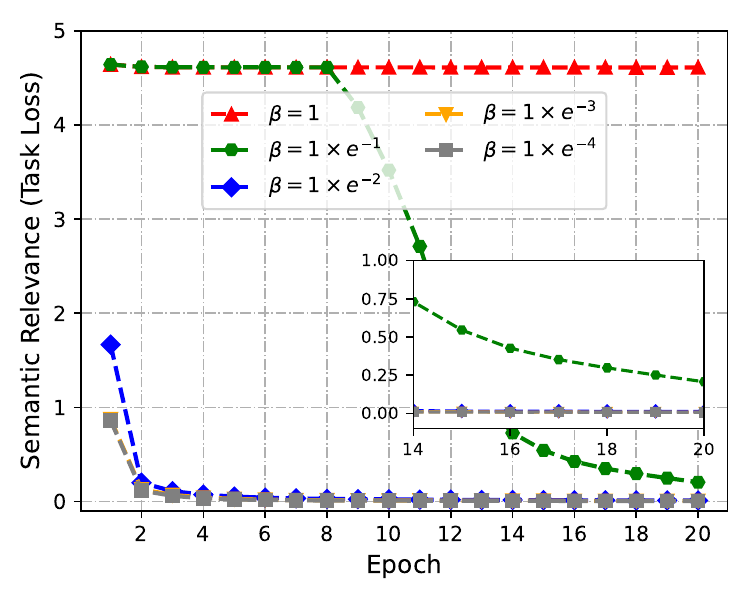}
        \label{fig:loss_beta_d}
    }
    \caption{Training loss curves of VDIB-JSCC with different $\beta \in  \{1, 10^{-1}, 10^{-2}, 10^{-3}, 10^{-4}\}$ at $\mathrm{PSNR}=15$ dB and $d_z = 512$. (a) and (c) depict the compression cost curves, while (b) and (d) present the semantic relevance curves.}
    \label{fig:loss_beta}
\end{figure}

\begin{figure}[!t]
    \centering
    \subfloat[Task Accuracy on CIFAR-10]{
        \includegraphics[width=0.47\linewidth]{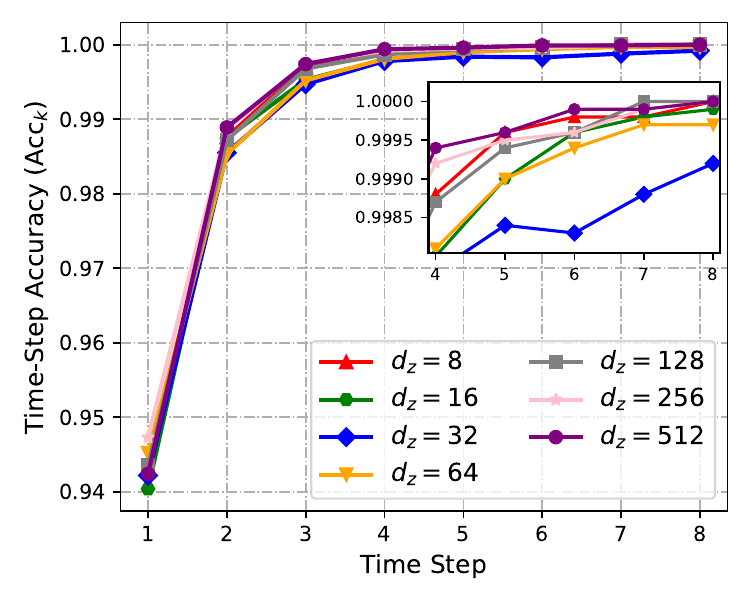}
        \label{fig:acc_dz_a}
    }
    \hfill
    \subfloat[Compression Cost on CIFAR-10]{
        \includegraphics[width=0.47\linewidth]{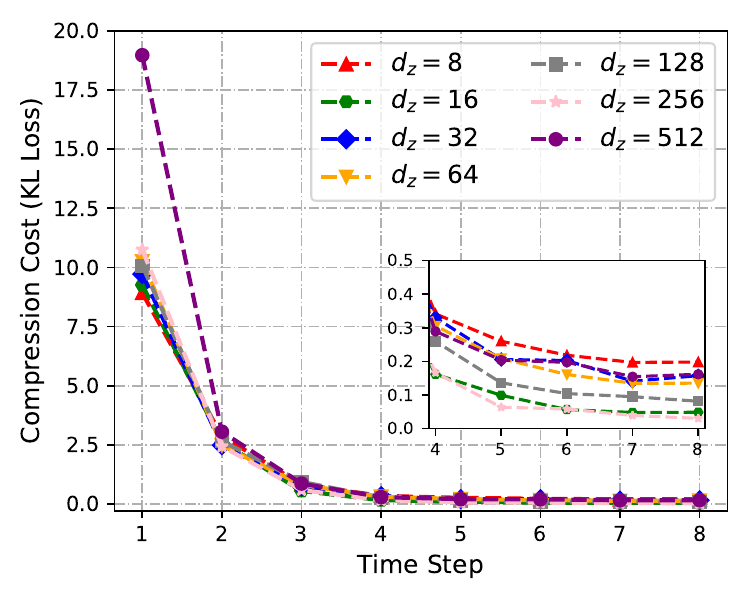}
        \label{fig:acc_dz_b}
    }
    \hfill
    \subfloat[Task Accuracy on CIFAR-100]{
        \includegraphics[width=0.47\linewidth]{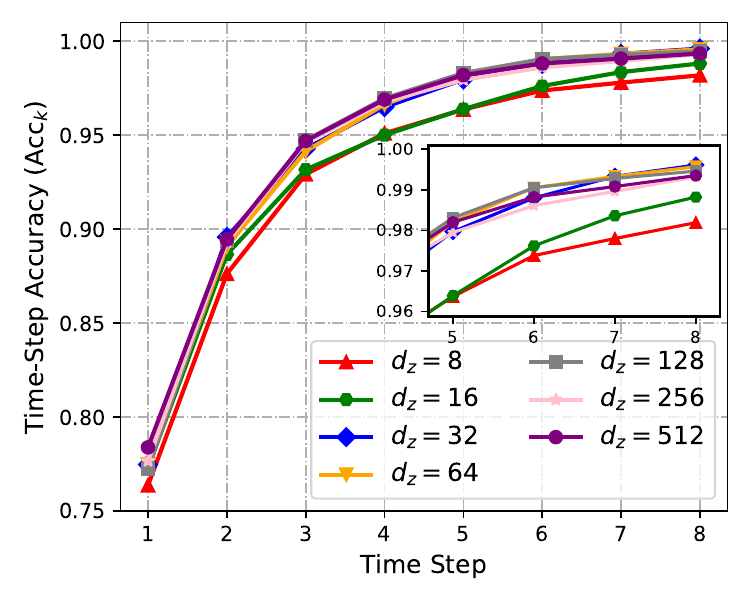}
        \label{fig:acc_dz_c}
    }
    \hfill
    \subfloat[Compression Cost on CIFAR-100]{
        \includegraphics[width=0.47\linewidth]{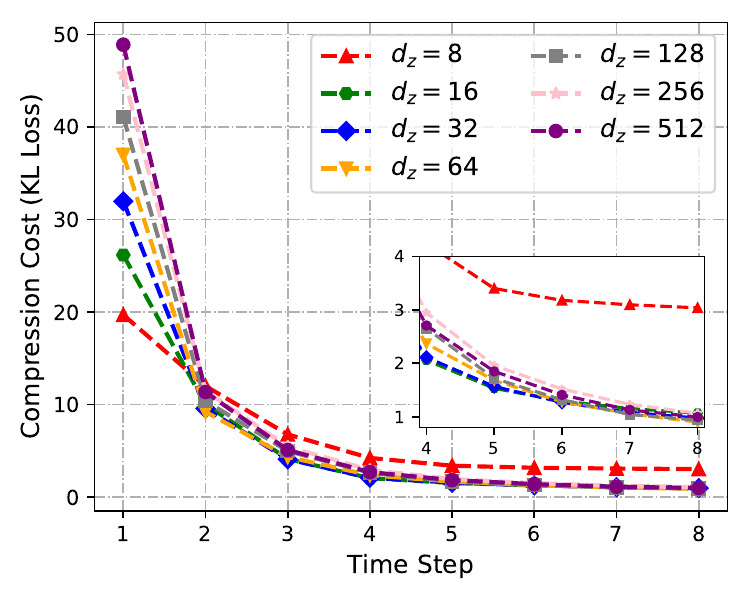}
        \label{fig:acc_dz_d}
    }
    \caption{Performance of VDIB-JSCC with different latent dimensions $d_z \in \{8, 16, 32, 64, 128, 256, 512\}$ at $\mathrm{PSNR}=15$ dB and $\beta=10^{-3}$. (a) and (c) show the time-step task accuracy, while (b) and (d) show the corresponding compression cost, i.e., the information regularization term in $\mathcal{L}_{\mathrm{VDIB}}$.}
    \label{fig:acc_dz}
\end{figure}

\subsection{Robustness}
We evaluate the effectiveness of the proposed VDIB-JSCC under different $\mathrm{PSNR}$ values, penalty parameter $\beta$, and communication latent dimensions $d_z$.

\subsubsection{Impact of different $\mathrm{PSNR}$ values under the AWGN channel}
Fig. \ref{fig:timeAcc_KLloss} and Fig. \ref{fig:time_step_acc} illustrate the performance of VDIB-JSCC under different $\mathrm{PSNR}$ values with the default setting.
Across different channel conditions, the task accuracy consistently increases rapidly in the early time steps and then improves only marginally, as shown in Figs. \ref{fig:timeAcc_kl_a} and \ref{fig:timeAcc_kl_c}.
For example, on CIFAR-100 with $\mathrm{PSNR} = 15$ dB, the task accuracy increases from $78.39\%$ at time step 1 to $94.70\%$ at time-step 3, corresponding to an increase of $16.31\%$, while a further increase of $4.65\%$ is observed from time-step 3 to time-step 8, reaching $99.35\%$.
Such a trend can be attributed to the substantial amount of task-relevant incremental information transmitted during the early time steps, with the amount of newly acquired incremental information gradually approaching zero as the time step increases. Consistent with this observation, the compression cost decreases as the time step increases, as shown in Figs. \ref{fig:timeAcc_kl_b} and \ref{fig:timeAcc_kl_d}.
For instance, on CIFAR-100 with $\mathrm{PSNR}=15$ dB, the compression cost decreases from $48.89$ at time-step 1 to $5.09$ at time-step 3 and further to $0.99$ at time-step 8.
Moreover, the proposed VDIB-JSCC is robust to variations in PSNR, as shown in Fig. \ref{fig:time_step_acc}, with the VDIB-JSCC trained on CIFAR-100 at $\mathrm{PSNR}=15$ dB achieving task accuracies ranging from $94.31\%$ to $94.36\%$ across different PSNR values.

\subsubsection{Impact of various penalty parameter $\beta$ on the DIB framework}
To verify Remark~\ref{rem:beta_DIB} on the selection of $\beta$ in the DIB framework, as shown in Fig. \ref{fig:timeAcc_beta_KLloss} and Fig. \ref{fig:loss_beta}, we evaluate VDIB-JSCC with different $\beta$ on CIFAR-10 and CIFAR-100.
Figs. \ref{fig:timeAcc_beta_kl_a} and \ref{fig:timeAcc_beta_kl_c} indicate that a smaller $\beta$ yields higher task accuracy, whereas an excessively large $\beta$ degrades the classification performance and eventually leads to task failure.
For instance, on CIFAR-100, the task accuracy increases from $77.92\%$ at time-step 1 to $99.31\%$ at time-step 8 for $\beta=10^{-4}$, whereas it remains at $10\%$ throughout all time steps for $\beta=1$.
Figs. \ref{fig:timeAcc_beta_kl_b} and \ref{fig:timeAcc_beta_kl_d} further show that an excessively large $\beta$ results in a very low compression cost at the initial time steps, corresponding to limited semantic information being transmitted initially.
For example, on CIFAR-100, the compression costs at time-steps 1 and 2 are $0.0019$ and $0.0015$, respectively, for $\beta=1$, with a total cost of $0.0034$. In contrast, the corresponding costs are $153.2708$ and $38.7274$ for $\beta=10^{-4}$, with a total cost of $191.9982$.
Those results verify the insight in Remark~\ref{rem:beta_DIB} that a smaller $\beta$ favors semantic preservation and retains more semantic information at a higher compression cost, whereas a larger $\beta$ retains less semantic information at a lower compression cost.
To further verify this observation, we present the training loss curves of VDIB-JSCC in Fig. \ref{fig:loss_beta}. The compression cost and semantic relevance are quantified by the information regularization term (referred to as KL loss) and task loss in $\mathcal{L}_{\mathrm{VDIB}}$, respectively. A smaller KL loss indicates less information encoded in the transmitted code, whereas a larger task loss indicates less task-relevant semantic information. As shown in Figs. \ref{fig:loss_beta_c} and \ref{fig:loss_beta_d}, on CIFAR-100, the KL loss converges to $0.0018$ for $\beta=1$ ($21.7177$ for $\beta=10^{-4}$), while the task loss converges to $4.6103$ for $\beta=1$ ($0.0059$ for $\beta=10^{-4}$). Thus, for $\beta=1$, VDIB-JSCC removes not only task-irrelevant redundant information but also discards task-relevant semantic information.

\subsubsection{Impact of varying communication latent dimension $d_z$}
Fig. \ref{fig:acc_dz} presents the performance of VDIB-JSCC with varying communication latent dimension $d_z$.
From Figs. \ref{fig:acc_dz_a} and \ref{fig:acc_dz_c}, we observe that the task accuracy on CIFAR-10 remains relatively stable across different $d_z$, indicating robustness to the communication latent dimension.
For instance, at time-step 4, the task accuracy ranges from $99.78\%$ to $99.94\%$  ($0.16\%$ difference) on CIFAR-10, and from $95.01\%$ to $96.96\%$ ($1.95\%$ difference) on CIFAR-100 across different $d_z$.
In contrast, a smaller $d_z$ degrades task performance on the more challenging CIFAR-100 task, with the task accuracy decreasing from $98.2\%$ for $d_z = 512$ to $96.37\%$ for $d_z = 8$ at time-step 5.
This result is consistent with our expectation, since a smaller $d_z$ provides insufficient representation capacity to encode all task-relevant semantic information.
Figs. \ref{fig:acc_dz_b} and \ref{fig:acc_dz_d} further support this finding by showing the corresponding compression costs.
On CIFAR-10, similar task accuracy across different $d_z$ values indicates that smaller $d_z$ values already provide sufficient task-relevant semantic information, while $d_z=512$ transmits substantially more redundant information, with a compression cost of $18.69$ at time-step 1 versus $10.74$ for $d_z=256$.
For the more challenging CIFAR-100 task, a smaller $d_z$ provides insufficient representation capacity to encode all task-relevant semantic information, with compression costs of $19.69$ and $3.04$ at time-steps 1 and 8, respectively, compared with $48.90$ and $0.99$ for $d_z=512$.

\section{Conclusion}\label{sec:conclusion}
In this paper, we have proposed a novel variational directed information bottleneck joint source-channel coding framework for task-oriented communication, termed VDIB-JSCC, enabling downstream tasks to be accomplished by transmitting only task-relevant incremental information.
VDIB-JSCC builds upon the directed information bottleneck framework by incorporating directed information into the information bottleneck, providing an information-theoretic formulation of task-oriented communication with incremental transmission.
Importantly, VDIB-JSCC demonstrates that retaining previously acquired knowledge can reduce the information required for downstream tasks, highlighting the role of historical knowledge base in communication-efficient task-oriented communication.
Furthermore, VDIB-JSCC can be adapted to agentic AI networks to exchange only task-relevant incremental information, thereby reducing communication load and facilitating efficient collaborative decision-making.
For the future work, we will investigate shared foundation models as parameterized knowledge base for many-to-many semantic communication among AI agents, leveraging shared knowledge to further reduce communication load.

\appendices
\section{Proof of the Theorem \ref{them:data_inequality}}\label{app:theorem_1}
To establish Eq. \eqref{eq:data_proc}, we first prove the following inequality chain:
\begin{equation}\label{eq:proof_chain_1}
\begin{aligned}
I(X^n \rightarrow \widehat{Z}^n)
&\le \sum_{k=1}^{n}
I(\widehat{Z}_k;Z_k|\widehat{Z}^{k-1})\\
&\le \sum_{k=1}^{n}
I(\widehat{Z}_k;Z_k)\\
&\le n\cdot C.
\end{aligned}
\end{equation}
Applying the chain rule of conditional mutual information to
\(I(X^k;Z^k,\widehat{Z}_k|\widehat{Z}^{k-1})\),
we obtain
\begin{equation}\label{eq:chain_mutual}
\begin{aligned}
&I(X^k;\widehat{Z}_k|\widehat{Z}^{k-1}) + I(X^k;Z^k|\widehat{Z}_k,\widehat{Z}^{k-1})\\
=& I(X^k;Z^k|\widehat{Z}^{k-1}) + I(X^k;\widehat{Z}_k|Z^k,\widehat{Z}^{k-1}).
\end{aligned}
\end{equation}
From the Markov relations of $X^k$, $Z^k$, and $\widehat{Z}^k$, as shown in Fig. \ref{fig:relationship}, we have $I(X^k;\widehat{Z}_k | Z^k, \widehat{Z}^{k-1}) = 0$.
Hence, 
\begin{equation}\label{eq:condi_mutu_info}
    I(X^k; \widehat{Z}_k | \widehat{Z}^{k-1}) \leq I(Z^k; \widehat{Z}_k | \widehat{Z}^{k-1}).
\end{equation}
where the inequality follows from the nonnegativity of conditional mutual information, i.e., $I(X^k;\widehat{Z}_k | \widehat{Z}_k, \widehat{Z}^{k-1}) \geq 0$.
Since $\widehat{Z}_k$ is conditionally independent of $Z^{k-1}$ given $\widehat{Z}^{k-1}$, as illustrated in Fig. \ref{fig:relationship}, we have
\begin{equation}\label{eq:equ_condi_mutu}
    I(Z^k; \widehat{Z}_k | \widehat{Z}^{k-1}) = I(Z_k; \widehat{Z}_k | \widehat{Z}^{k-1}). 
\end{equation}
Combining Eq. \eqref{eq:condi_mutu_info} and Eq. \eqref{eq:equ_condi_mutu} yields
\begin{equation}
    I(X^k; \widehat{Z}_k | \widehat{Z}^{k-1}) \leq I(Z_k; \widehat{Z}_k | \widehat{Z}^{k-1}),
\end{equation}
Summing over \(k = 1,\dots,n\), we obtain
\begin{equation}
    \sum_{1 \leq k \leq n} I(X^k; \widehat{Z}_k | \widehat{Z}^{k-1}) \leq \sum_{1 \leq k \leq n}  I(Z_k; \widehat{Z}_k | \widehat{Z}^{k-1}).
\end{equation}
By definition of directed information (detailed in Definition \ref{def:directed_information}), we have
\begin{equation}\label{eq:conclu_1}
    I (X^n \rightarrow \widehat{Z}^n) \leq \sum_{1 \leq k \leq n}  I(Z_k; \widehat{Z}_k | \widehat{Z}^{k-1}).
\end{equation}
In addition, since conditioning cannot increase mutual information, we have
\begin{equation}\label{eq:conclu_2}
    \sum_{1 \leq k \leq n} I(Z_k; \widehat{Z}_k | \widehat{Z}^{k-1}) \leq \sum_{1 \leq k \leq n} I(Z_k;\widehat{Z}_k).
\end{equation}
Moreover, noting that $C = \max_{k \in [1,n]} I(\widehat{Z}_k;Z_k)$, we further obtain
\begin{equation}\label{eq:conclu_3}
    \sum_{1 \leq k \leq n} I(Z_k;\widehat{Z}_k | \widehat{Z}^{k-1}) \leq n \cdot C.
\end{equation}
As a result, Eq. \eqref{eq:proof_chain_1} follows from Eqs. \eqref{eq:conclu_1}–\eqref{eq:conclu_3}.
Next, we prove that inequality $I(\widehat{Z}^n \rightarrow Y^n) \leq I(X^n \rightarrow \widehat{Z}^n)$ holds.
Specifically, we apply the chain rule of conditional mutual information to $I(Y^k, X^k; \widehat{Z}_k | \widehat{Z}^{k-1})$, yielding
\begin{equation}
\begin{aligned}
    &I(X^k;\widehat{Z}_k | \widehat{Z}^{k-1}) + I(Y^k; \widehat{Z}_k | X^k, \widehat{Z}^{k-1}) \\
    =& I(Y^k; \widehat{Z}_k | \widehat{Z}^{k-1}) + I(X^k; \widehat{Z}_k | Y^k, \widehat{Z}^{k-1}).
\end{aligned}
\end{equation}
By the Markov relations among $Y^k$, $X^k$, and $\widehat{Z}^k$, as shown in Fig. \ref{fig:relationship}, it holds that $I(Y^k;\widehat{Z}_k | X^k, \widehat{Z}^{k-1}) = 0$. Thus,
\begin{equation}
    I(Y^k;\widehat{Z}_k | \widehat{Z}^{k-1}) \leq I(X^k; \widehat{Z}_k | \widehat{Z}^{k-1}). 
\end{equation}
Summing over $k=1, \cdots, n$, we obtain
\begin{equation}
    \sum_{1 \leq k \leq n}  I(Y^k;\widehat{Z}_k | \widehat{Z}^{k-1}) \leq \sum_{1 \leq k \leq n} I(X^k; \widehat{Z}_k | \widehat{Z}^{k-1}). 
\end{equation}
According to the definition of directed information (detailed in Definition \ref{def:directed_information}), it follows that 
\begin{equation}
    I(Y^n \rightarrow \widehat{Z}^n) \leq I(X^n \rightarrow \widehat{Z}^n).
\end{equation}
In summary, Eq. \eqref{eq:data_proc} in Theorem \ref{them:data_inequality} is proved.
In particular, when considering the average information flow, Eq. \eqref{eq:them_2} in Theorem \ref{them:data_inequality} holds.

\section{Proof of the Theorem \ref{them:bound_DIB}}\label{app:theorem_2}
We first consider the conditional entropy term $H(\widehat{Z}_k| \widehat{Z}^{k-1})$ appearing in Eq.~\eqref{eq:entropy_first}, which can be expressed as
\begin{equation}
H(\widehat{Z}_k \mid \widehat{Z}^{k-1})=-\mathbb{E}_{P(\widehat{z}^k)}\left[\log P(\widehat{z}_k \mid \widehat{z}^{k-1})\right].
\end{equation}
However, the conditional prior distribution $P(\widehat{z}_k \mid \widehat{z}^{k-1})$ is generally intractable. To address this issue, we introduce an autoregressive prior $r_{\psi} (\widehat{z}_k | \widehat{z}^{k-1})$ to approximate it. By the non-negativity of the KL-divergence, we have
\begin{equation}
    D_{\mathrm{KL}}(P(\widehat{z}_k | \widehat{z}^{k-1}) \| r_{\psi}(\widehat{z}_k | \widehat{z}^{k-1})) \geq 0
\end{equation}
which implies
\begin{equation}
    \mathbb{E}_{P(\widehat{z}^k)}[\log P(\widehat{z}_k | \widehat{z}^{k-1})] \geq \mathbb{E}_{P(\widehat{z}^k)}[\log r_{\psi}(\widehat{z}_k | \widehat{z}^{k-1})].
\end{equation}
Consequently, the conditional entropy admits the following upper bound
\begin{equation}
    H(\widehat{Z}_k | \widehat{Z}^{k-1}) \leq - \mathbb{E}_{P(\widehat{z}^k)}[\log r_{\psi}(\widehat{z}_k | \widehat{z}^{k-1})].
\end{equation}
Similarly, for the conditional entropy $H(\widehat{Z}_k \mid X^k, \widehat{Z}^{k-1})$ appearing in Eq.~\eqref{eq:entropy_third}, we introduce a semantic encoder $\mathcal{A}_{\theta}(\widehat{z}_k \mid x^k, \widehat{z}^{k-1})$ to approximate the true posterior $P(\widehat{z}_k \mid x^k, \widehat{z}^{k-1})$.
By the non-negativity of the KL-divergence, i.e.,
\begin{equation}
    D_{\mathrm{KL}}(P(\widehat{z}_k | x^k, \widehat{z}^{k-1}) \| \mathcal{A}_{\theta}(\widehat{z}_k | x^k, \widehat{z}^{k-1})) \geq 0,
\end{equation}
we obtain
\begin{equation}
    H(\widehat{Z}_k| X^k, \widehat{Z}^{k-1}) \leq - \mathbb{E}_{P(y^k,\widehat{z}^k)} \left[\log \mathcal{A}_{\theta}(\widehat{z}_k | x^k, \widehat{z}^{k-1}) \right].
\end{equation}
Next, we consider the posterior distribution $P(\widehat{z}_k \mid y^k, \widehat{z}^{k-1})$ appearing in the conditional entropy $H(\widehat{Z}_k | Y^k, \widehat{Z}^{k-1})$ (i.e., Eq.~\eqref{eq:entropy_second}).
Using the conditional independence relations 
$\widehat{Z}_k \perp\!\!\!\perp Y^{k-1} | \widehat{Z}^{k-1}$ (i.e., the current semantic embedding is generated solely from the historical semantic embeddings, without access to the historical targets) and $Y_k \perp\!\!\!\perp Y^{k-1} | \widehat{Z}^k$ (i.e., the semantic embedding sequence $\widehat{Z}^k$ provides a sufficient temporal information for predicting the current target), it follows that
\begin{align}
P(\widehat{z}_k | y^{k-1}, \widehat{z}^{k-1}) &= P(\widehat{z}_k | \widehat{z}^{k-1}), \\
P(y_k | y^{k-1}, \widehat{z}^{k-1}, \widehat{z}_k) & = P(y_k | \widehat{z}^k).
\end{align}
Applying Bayes' rule, we obtain 
\begin{equation}
P(\widehat{z}_k \mid y^k, \widehat{z}^{k-1}) = \frac{ P(y_k \mid \widehat{z}^k)\,
P(\widehat{z}_k \mid \widehat{z}^{k-1}) }{ P(y_k \mid y^{k-1}, \widehat{z}^{k-1})}.
\end{equation}
Hence, we have the proportional form
\begin{equation}
P(\widehat{z}_k \mid y^k, \widehat{z}^{k-1}) \propto P(y_k \mid \widehat{z}^k) P(\widehat{z}_k \mid \widehat{z}^{k-1}),
\end{equation}
since the proportionality follows because the normalizing factor $P(y_k | y^{k-1}, \widehat{z}^{k-1})$ is independent of $\widehat{z}_k$.
To facilitate tractable inference, we introduce a variational decoder
$\mathcal{B}_{\phi}(y_k \mid \widehat{z}^k)$
to approximate $P(y_k \mid \widehat{z}^k)$, and combine it with the prior approximation $r_{\psi}(\widehat{z}_k \mid \widehat{z}^{k-1})$.
By the non-negativity of the KL-divergence, the conditional entropy satisfies
\begin{equation}
\begin{aligned}
H(\widehat{Z}_k \mid Y^k, \widehat{Z}^{k-1}) &\leq -\mathbb{E}_{P(y^k,\widehat{z}^k)} \left[ \log \mathcal{B}_{\phi}(y_k \mid \widehat{z}^k) \right] \\
-&\mathbb{E}_{P(y^k,\widehat{z}^k)} \left[ \log r_{\psi}(\widehat{z}_k \mid \widehat{z}^{k-1}) \right].
\end{aligned}
\end{equation}
Using the causal encoder factorization
\begin{equation}
    \mathcal{A}_{\theta}(\widehat{z}^k \| x^k) = \prod_{t=1}^k \mathcal{A}_{\theta}(\widehat{z}_t | x^t, \widehat{z}^{t-1}),
\end{equation}
we have
\begin{align}
    P(\widehat{z}^k) &= \int P(x^k) \mathcal{A}_{\theta}(\widehat{z}^k \| x^k) dx^k, \\
    P(y^k, \widehat{z}^k) & = \int P(x^k,y^k) \mathcal{A}_{\theta}(\widehat{z}^k \| x^k) dx^k.
\end{align}
Consequently,
\begin{align}
    &H(\widehat{Z}_k | \widehat{Z}^{k-1}) \nonumber \\
    \leq& -\mathbb{E}_{P(x^k)} \left\{ \mathbb{E}_{\mathcal{A}_{\theta}(\widehat{z}^k \| x^k)} \left[  \log r_{\psi}(\widehat{z}_k | \widehat{z}^{k-1})  \right] \right\}, \\
    &H(\widehat{Z}_k | X^k, \widehat{Z}^{k-1}) \nonumber \\
    \leq & -\mathbb{E}_{P(x^k,y^k)} \left\{ \mathbb{E}_{\mathcal{A}_{\theta}(\widehat{z}^k \| x^k)} \left[ \log \mathcal{A}_{\theta}(\widehat{z}_k | x^k, \widehat{z}^{k-1}) \right]  \right\}, \\
    &H(\widehat{Z}_k | Y^k, \widehat{Z}^{k-1}) \nonumber \\
    \leq & -\mathbb{E}_{P(x^k,y^k)} \left\{ \mathbb{E}_{\mathcal{A}_{\theta}(\widehat{z}^k \| x^k)} \left[ \log \mathcal{B}_{\phi}(y_k | \widehat{z}^k) \right]  \right\} \nonumber\\
    & -\mathbb{E}_{P(x^k,y^k)} \left\{ \mathbb{E}_{\mathcal{A}_{\theta}(\widehat{z}^k \| x^k)} \left[ \log r_{\psi}(\widehat{z}^k | \widehat{z}^{k-1}) \right]  \right\}.
\end{align}
Finally, substituting the above variational bounds into the entropy representation of the DIB objective yields
\begin{equation}
\begin{aligned}
    &\mathcal{L}_{\mathrm{DIB}} \leq \sum_{k=1}^n \Big\{ \mathbb{E}_{P(x^k,y^k)} \{ \mathbb{E}_{\mathcal{A}_{\theta}(\widehat{z}^k \| x^k)} [- \log \mathcal{B}_{\phi}(y_k | \widehat{z}^k) ] \} \\
    &+\beta \underbrace{\mathbb{E}_{P(x^k,y^k)} \{ \mathbb{E}_{\mathcal{A}_{\theta}(\widehat{z}^k \| x^k)} [\log \frac{\mathcal{A}_{\theta}(\widehat{z}_k \mid x^k, \widehat{z}^{k-1})}{r_{\psi}(\widehat{z}_k \mid \widehat{z}^{k-1})} ] \} }_{D_{\mathrm{KL}} ( \mathcal{A}_{\theta}(\widehat{z}_k \mid x^k, \widehat{z}^{k-1}) \| r_{\psi}(\widehat{z}_k \mid \widehat{z}^{k-1})  )} \Big\},
\end{aligned}
\end{equation}
where the right-hand side is precisely the objective  $\mathcal{L}_{\mathrm{VDIB}}$  defined in the Theorem \ref{them:bound_DIB} (i.e., Eq. \eqref{eq:VDIB}).
This completes the proof of Theorem \ref{them:bound_DIB}.

\bibliographystyle{IEEEtran}
\bibliography{sample-base}




\vspace{11pt}




\vfill

\end{document}